\documentclass[manuscript,screen,nonacm]{acmart}
\AtBeginDocument{%
  }

\begin{document}

\title{Exploring the Temporal Design of Olfactory–Auditory Takeover Reminders in SAE Level 3 Automated Driving}

\author{Jiuyi Wang}
 \affiliation{
  \department{School of Computer Science and Engineering}
  \institution{University of Electronic Science and Technology of China}
  \city{Chengdu}
  \country{China}
 }
\author{Liang Luo}
 \affiliation{
  \department{School of Computer Science and Engineering}
  \institution{University of Electronic Science and Technology of China}
  \city{Chengdu}
  \country{China}
 }
\author{Yu Gu}
 \affiliation{
  \department{School of Computer Science and Engineering}
  \institution{University of Electronic Science and Technology of China}
  \city{Chengdu}
  \country{China}
 }
\author{Fuji Ren}
 \affiliation{
  \department{School of Computer Science and Engineering}
  \institution{University of Electronic Science and Technology of China}
  \city{Chengdu}
  \country{China}
 }

\begin{abstract}
Conditional automated driving requires drivers to resume control after monitoring. We examine olfactory--auditory takeover reminders as a temporal evidence problem: software cueing, manual actuation, odor arrival, and perceptual detection need not coincide. We developed a CARLA-based instrument and conducted an exploratory study using three arrangements: audio only, a scent-release cue aligned with the auditory takeover request (TOR), and an earlier scent-release cue. Technical interruptions left 33 analyzable TOR trials from 18 participants. Retrospective odor detection was reported in 8/13 aligned observations but only 2/8 earlier-cue observations, while odor arrival and perceptual onset were unmeasured. Behavioral, safety, and subjective measures showed divergent patterns. We contribute an event-level temporal representation, a five-step evidence-accounting workflow, and replication requirements for latency-bearing multimodal channels. Behavioral and interview patterns are treated as hypothesis-generating rather than evidence for optimal timing, perceptual ordering, or cognitive mechanisms.
\end{abstract}



\begin{teaserfigure}
 \includegraphics[width=\textwidth]{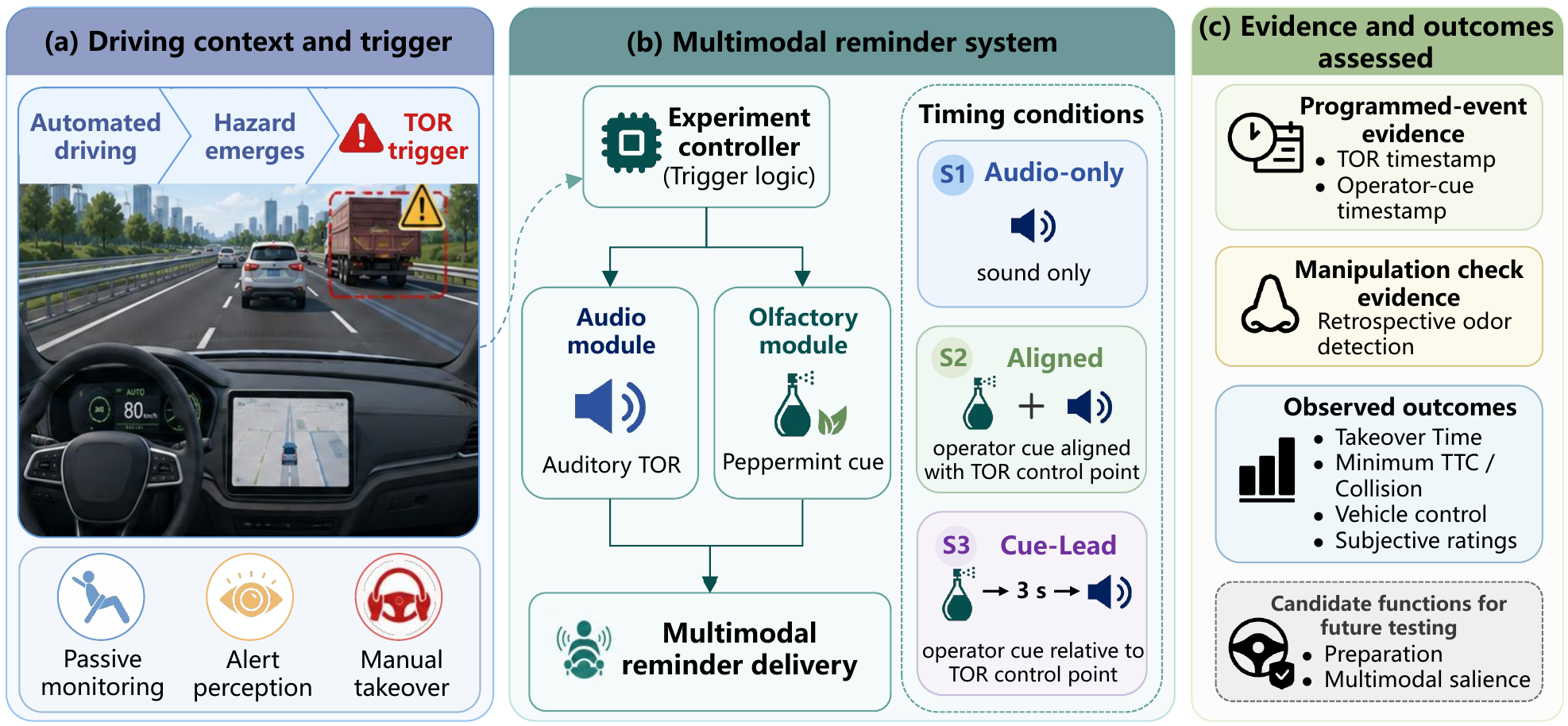}
  \caption{Overview of the research framing. Audio-only retains the explicit auditory TOR; Aligned (S2) adds an olfactory operator release cue at the auditory TOR control point; Cue-Lead (S3) places the same olfactory operator cue 3.0 s earlier. The study evaluates behavioral, safety, vehicle-control, and experience outcomes rather than ranking modalities by reaction time alone.}
  \Description{Overview of the multimodal takeover reminder framework.}
  \label{fig:overview}
\end{teaserfigure}

\maketitle

\section{Introduction}
\subsection{Background and Problem}
SAE Level 3 conditional automation performs the dynamic driving task within a defined operational design domain while requiring a fallback-ready user to resume control when requested \cite{on2021taxonomy}. This changes the driver's role from continuous perception--action coupling to supervisory monitoring, making the return to manual control a safety-critical transition rather than a simple notification event \cite{parasuraman1997humans,endsley1995out,endsley1995toward,parasuraman2000model,lee2004trust,chen2025systematic}. A takeover request (TOR) must be noticed and interpreted, after which the driver must reorient to the road, select an action, intervene, and stabilize the vehicle \cite{zhang2019determinants,liang2026towards,wu2025request,tan2026there,zeeb2015determines,merat2014transition,vlakveld2018situation,heinrich2025ready,louw2017coming}.

This transition spans several stages: a reminder may attract attention yet leave insufficient time to reconstruct the situation, while a rapid control input may still produce a poorly coordinated maneuver. From an HCI perspective, the reminder mediates a transfer of control between automation and driver, so evaluation should capture both re-entry and subsequent maneuver quality rather than first input alone.

Takeover time is therefore informative but incomplete. Faster responses can coexist with smaller safety margins, abrupt braking, or substantial lateral movement \cite{zeeb2016take,jin2025impact,louw2017coming,weaver2022systematic}. Prolonged low-engagement automated driving can further increase sleepiness and reduce readiness through passive fatigue \cite{may2009driver,saxby2007development,matthews2002task,thiffault2003monotony,pan2025exploring,coyne2026effect,zhang2026novel,wang2025exploring}. Our protocol accordingly used 20 min of monotonous automated driving with basic road monitoring and KSS state checks \cite{aakerstedt1990subjective,kaida2006validation}. We treat this as a passive-fatigue context rather than evidence of a uniform physiological state.

We intentionally omitted a non-driving task to isolate low-load supervisory monitoring and reduce task-specific interference. This choice narrows generalizability to one controlled readiness context rather than the broader range of visually or cognitively demanding Level 3 situations.

\subsection{Research Gap: From Multimodality to Temporal Semantics}
Takeover research has extensively examined lead time, traffic criticality, non-driving tasks, driver state, trust, and warning modality \cite{zhang2019determinants,wu2025effect,dong2026multimodal,srour2025all,wang2024human}. Multimodal TORs are often described through modality combinations or nominal onset times, implicitly treating channel timing as directly comparable \cite{hong2022effect,yu2025impact,zhang2025effects,shi2026striving}. That assumption becomes problematic for modalities whose output is delayed, spatially transported, persistent, or only indirectly coupled to a software trigger.

Olfactory takeover support itself is not new. Most directly, Tang et al. paired peppermint stimulation with auditory and tactile takeover requests in a 60-participant highly automated driving study involving awakening from light sleep \cite{tang2021olfactory}.Our gap is therefore not the use of scent itself, but how a slower olfactory channel should be represented and interpreted when programmed timing differs from experienced timing. An olfactory event may involve software cueing, device actuation, physical transport or arrival, and perceptual detection \cite{obrist2014opportunities,spence2017digitizing,maggioni2020smell,liu2024intelligent,dmitrenko2016comparison}. In our apparatus, manual experimenter actuation adds another temporal link, so a programmed alignment or lead does not establish perceptual simultaneity or precedence.

\subsection{Research Questions and Contributions}
The study addresses two research questions:

\noindent\textbf{RQ1.} Under monotonous, passive-fatigue conditions, what behavioral, safety, and vehicle-control patterns are observed across three programmed reminder arrangements: Audio-only (S1), TOR-Control-Point Aligned (S2), and a $-3$ s Operator-Cue Lead (S3)?

\noindent\textbf{RQ2.} How do subjective ratings, retrospective odor-detection responses, and interview accounts align with---or diverge from---the programmed temporal relations, and what does this imply for interpreting slow multimodal cues?

We position the paper as a formative HCI methodology study rather than a confirmatory test of condition effects. Its primary contribution is temporal evidence accounting: explicitly separating the events the experimental system controls from physical and perceptual events it does not measure, and then constraining claims to the evidence available at each stage. The empirical comparison serves as a demonstration of why this distinction matters.

This paper makes three contributions. First, we articulate an event-level representation for olfactory--auditory TOR timing that distinguishes software cueing, manual actuation, physical transport/arrival, perceptual detection, and the fixed auditory TOR reference. Second, we operationalize this representation as a five-step temporal evidence-accounting workflow (Sec.~\ref{sec:evidence-accounting-workflow}) and instantiate it in a CARLA-based research instrument that keeps controlled, logged, reported, and unobserved events distinguishable. Third, we use an incomplete exploratory dataset to demonstrate the interpretation consequences of this accounting: programmed alignment or lead cannot be translated directly into perceptual simultaneity or precedence, and behavioral, safety, and experience measures can diverge without establishing a causal timing effect.

The empirical component contains 33 of 54 planned condition slots from 18 participants. Because technical interruptions disrupted the intended repeated-measures design and condition was bundled with scenario, we treat condition-level outcome patterns as descriptive demonstration evidence rather than a ranking of timing strategies. The detection audit and the resulting interpretation boundaries are therefore part of the paper's methodological result, and they motivate concrete requirements for a fully crossed replication with instrumented scent delivery. The overall framing is summarized in Fig.~\ref{fig:overview}.

\section{Related Work}
\subsection{Takeover in Automated Driving}
Conditional automation creates an out-of-the-loop transition problem: drivers who are relieved from continuous control must rapidly reconstruct the driving situation and resume motor control when automation can no longer continue \cite{endsley1995out,lee2004trust,chen2025systematic}. Meta-analytic and simulator studies show that takeover performance depends on multiple interacting factors, including warning lead time, non-driving activities, driver state, traffic context, and warning modality \cite{zhang2019determinants,weaver2022systematic,hu2024non,li2025obstacle,morgan2017manual,guo2024analysis}. Takeover time is therefore only one stage of the transition. Studies distinguishing first input from takeover quality show that rapid responses can coexist with poorer lateral or longitudinal control, motivating evaluation across timing, safety margin, and maneuver behavior \cite{zeeb2016take,jin2025impact,louw2017coming}.

This multidimensional view also affects how warning designs should be compared. A design that reduces response latency may still impose abrupt braking, encourage a different steering strategy, or alter subjective confidence in ways that are invisible in a single timing metric. Reviews of takeover research increasingly emphasize that traffic context and measurement choice shape apparent conclusions \cite{weaver2022systematic,deng2024design}. We therefore treat behavioral, safety, control, and experience measures as complementary rather than interchangeable indicators.

Driver readiness also changes over time. Prolonged automated driving can increase subjective sleepiness and reduce engagement, particularly in monotonous settings \cite{pan2023impact,yin2024monitoring,coyne2026effect,loew2025impact,pan2024age,ma2026video,liu2024safety,sun2026adaptive,sheem2026must,zhang2025objective}. KSS provides a validated subjective sleepiness measure \cite{aakerstedt1990subjective,kaida2006validation}, but it does not directly measure visual attention, situation awareness, or motor preparation at TOR onset.

Trust and prior experience further complicate readiness. Drivers may respond differently when they expect automation to be reliable, when they are familiar with a particular TOR, or when a cue has acquired meaning through repeated exposure \cite{singh2025trust,kaufman2025did,alkurdi2025trust}. These factors are especially relevant for an unfamiliar olfactory cue: an advance scent may be interpreted as preparatory only after participants have learned that it is related to a takeover sequence. Our design therefore preserves an explicit auditory command rather than relying on learned scent semantics alone. This distinction matters for our interpretation: behavioral response and interview accounts can suggest preparatory effects without establishing a specific attentional mechanism.

\subsection{Multimodal Human--Vehicle Interaction}
Auditory TORs are widely used because they do not require a driver to look at a display and can deliver an explicit command quickly. Visual, speech, haptic, and multimodal interfaces can add redundancy, contextual information, direction, or urgency \cite{dong2026multimodal,srour2025all,hong2022effect,rodak2025hmi,srour2024s,yu2025impact,chai2024effects,huang2022takeover,bazilinskyy2018take}. However, salience and user experience need not move together with safety outcomes. Recent work therefore increasingly examines not only whether a reminder is detected but how modality, urgency, trust, workload, and acceptance jointly shape the transition \cite{wang2024human,shi2026striving,kaufman2025did}.

A useful design distinction is between a channel that carries the explicit command and a channel that provides contextual or preparatory support. Multimodal systems do not require every channel to carry the same message at the same time. Prior ambient and preparatory interfaces suggest that secondary channels can be used to orient a user before a more explicit instruction \cite{sadeghian2018feel,tan2025driver,zhang2025actual,van2016supporting,walch2017car,wintersberger2018let}. Our reminder architecture adopts this role separation: sound always communicates the takeover demand, while scent is allowed to vary temporally without bearing the full semantic burden of the safety-critical instruction.

Olfactory HCI contributes a different set of constraints. Odors are ambient and can persist beyond the trigger event; their delivery depends on hardware, airflow, concentration, individual sensitivity, and learned associations \cite{obrist2014opportunities,brewster2006olfoto,fu2024aromablendz,spence2017digitizing,maggioni2020smell,chang2025crossmodal,wang2025multi,obrist2026follow}. Prior work has explored scent-delivery devices, driving-related message mappings, multi-scent notification, and automotive applications \cite{dmitrenko2016comparison,dmitrenko2017did,dmitrenko2018smell,dmitrenko2020caroma,schartmuller2020sick}. These studies establish that in-car olfactory interaction is feasible, but also highlight residual odor, preference variation, and device latency.

Olfaction also differs from transient visual or auditory output in persistence and personalization. The same concentration can be experienced differently across people, cabin airflow can alter delivery, and residual scent can affect later trials or repeated-use acceptance \cite{dmitrenko2016comparison,maggioni2020smell,brooks2023third}. These properties make washout, device placement, and timing semantics part of the interaction design itself, not merely implementation details. We therefore use peppermint as an auxiliary cue rather than a stand-alone safety command and do not interpret it as a proven fatigue countermeasure.

\subsection{Timing, Multi-Stage Warnings, and Olfactory Coordination}
\label{sec:timing-coordination}
Warning lead time is a central determinant of takeover opportunity: too little time compresses perception and action, whereas very early warnings can reduce urgency or become easy to ignore \cite{zhang2019determinants,liang2026towards,wu2025request,wu2025effect}. Recent two-stage and multi-stage warning research explicitly separates a preparatory stage from a later takeover command. For example, work on two-stage intervals and modality combinations asks how much advance notice should precede an explicit request, while multi-stage TOR designs examine how preparatory information can support the transition when immediate takeover is not required \cite{zhang2025effects,tan2026there,wu2025request}. This literature is closely related to our temporal framing.

It also provides a useful conceptual baseline: a two-stage warning assumes that an earlier event has a different functional role from the final TOR. The first stage may signal an approaching transition, create time for reorientation, or communicate context, while the later stage preserves an unambiguous request for action. Our S3 $-3$ s Operator-Cue Lead arrangement follows this general two-stage structure at the level of programmed control events, but the earlier event is generated through a modality whose delivery latency is itself uncertain.

Our study differs in the temporal semantics of the first stage. In many visual, auditory, or haptic two-stage designs, software onset and physical stimulus availability are more tightly coupled, so a nominal 3-s interval can more closely approximate the delivered stimulus interval. For olfaction, the same assumption is not defensible. A software release cue precedes manual actuation, physical emission, transport through air, and detection, and scent can persist after the event \cite{liu2024intelligent,spence2017digitizing,dmitrenko2016comparison}. Thus, an olfactory 3-s lead is not another instance of a precisely timed first warning. It requires the experimental system to distinguish controllable software timing from unmeasured physical and perceptual timing.

This distinction defines our incremental contribution relative to prior multi-stage warning work. We do not claim to identify an optimal two-stage interval. Instead, we study how a slower auxiliary modality can be positioned relative to an explicit auditory TOR while preserving a common analysis origin, and we make the cue-to-actuation uncertainty part of the system model rather than hiding it. The pilot-selected 3-s offset is therefore one calibrated point in a larger temporal-design space. Replications should vary both the interval and hazard time budget, ideally with automated scent actuation and measured odor concentration near the driver.

\section{System Design}
\subsection{System Overview}
The system was built as a research instrument for manipulating and auditing multimodal reminder timing rather than as a production vehicle interface. It had three design goals: (1) expose the software-level timing relation between auxiliary olfactory cueing and the explicit auditory TOR, (2) keep the auditory TOR semantically stable across arrangements, and (3) log the controlled software events, participant input, and vehicle response while explicitly representing manual actuation, physical odor arrival, and perceptual detection as distinct links with different evidence status.

The framework integrates CARLA scenario execution, steering-wheel and pedal input, auditory playback, an experimenter-facing Operator Panel, manual scent delivery, questionnaires, and synchronized telemetry/event logs.

\begin{figure}[h]
  \centering
 \includegraphics[width=\linewidth]{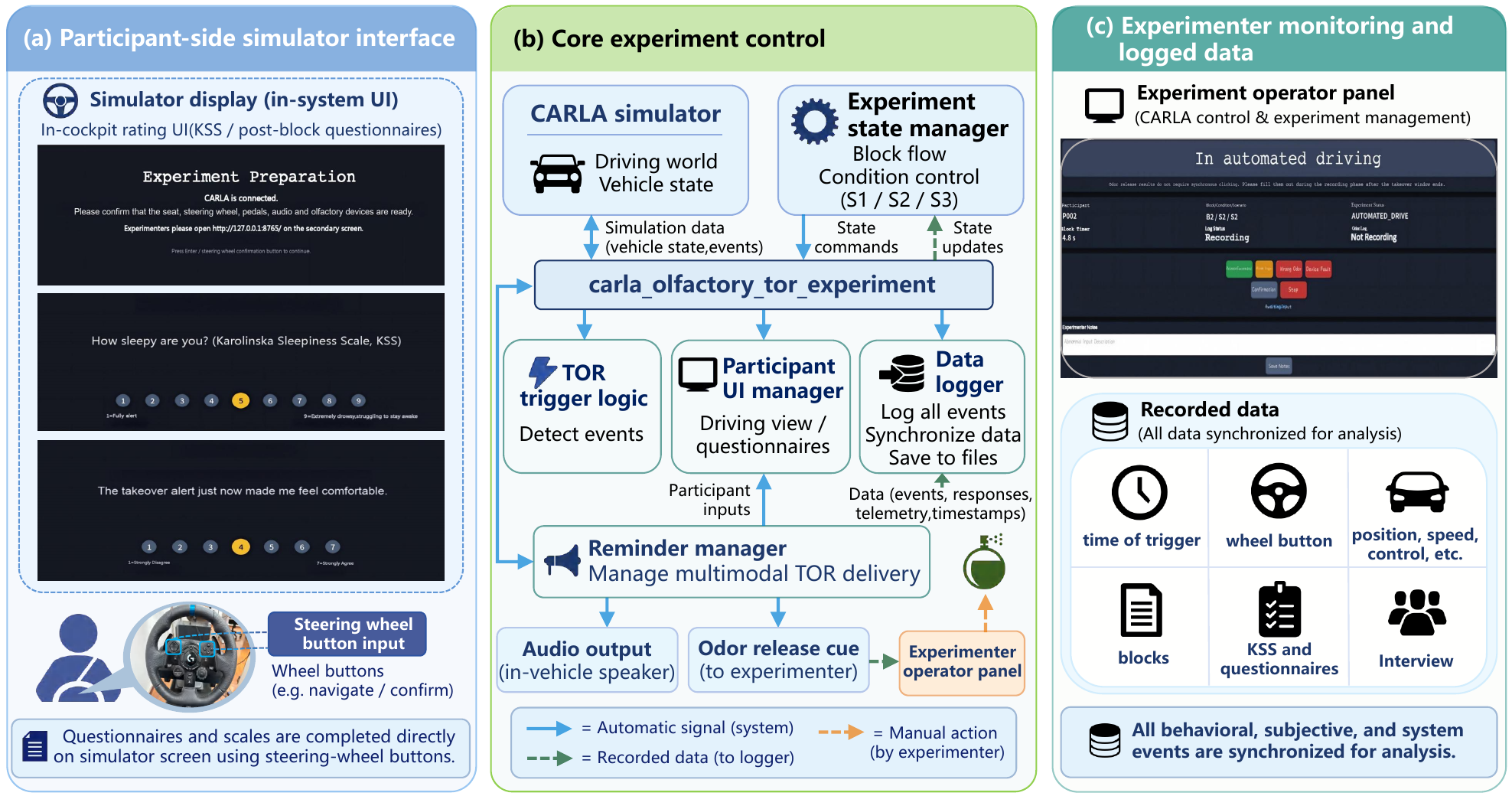}
  \caption{System architecture of the CARLA-based experimental framework, integrating scenario execution, auditory TOR playback, an experimenter-facing Operator Panel, manual olfactory delivery, steering-wheel and pedal input, participant questionnaires, and synchronized telemetry/event logging.}
  \Description{System architecture.}
  \label{fig:architecture}
\end{figure}

The Operator Panel was used because the scent device was not directly software-actuated in the formal setup. It exposed condition state and the moment at which the experimenter should operate the scent device while remaining outside the participant's view. This arrangement allowed the study to manipulate the timing of the operator instruction even though physical emission remained manual. It also created a clear evidence boundary: the system can timestamp the software cue, but it cannot infer the exact moment at which odor reached or was noticed by the participant. CARLA was selected for programmable road geometry and deterministic simulation control \cite{dosovitskiy2017carla}. Simulator studies offer repeatability and safety but trade off ecological realism \cite{kaptein1996driving,li2025researchers}. As shown in Fig.~\ref{fig:architecture}, the experimental controller links cue presentation, participant input, and vehicle state within a common workflow.

\subsection{Multimodal Reminder Mechanism}
\subsubsection{Modality Roles and Auditory Standardization}
The reminder design used a stable primary--secondary relationship. The auditory TOR was the explicit action request in every arrangement; the olfactory channel was an auxiliary cue whose presence and software/operator timing varied. We therefore name the arrangements according to the events under experimental control: Audio-only (S1) omitted scent; TOR-Control-Point Aligned (S2) issued the experimenter-facing scent-release cue at the auditory TOR software-control point; and the $-3$ s Operator-Cue Lead (S3) issued the same cue 3.0 s before the auditory TOR and did not issue a second scent cue at TOR. These labels intentionally avoid claiming perceptual simultaneity or perceptual precedence.

At the reminder-programming level, S2 and S3 used the same auditory signal and the same olfactory delivery pathway, with the programmed operator-cue relation as the intended difference. However, the formal protocol also bound reminder arrangement to predefined scenario bundles (Sec.~\ref{sec:conditions}), so the realized comparison should not be interpreted as isolating a causal timing effect. The auditory stimulus itself was fixed across conditions: the same \texttt{tor\_beep.wav} file (mono, 16-bit PCM, 44.1 kHz, 0.800 s) was played at software volume 0.75.

\subsubsection{Pilot Calibration}
A pre-study pilot with eight participants (age 20--26, $M=23.75$) was used to select a detectable, tolerable scent configuration and calibrate the operator-cue interval. The formal mixture used 20 drops of peppermint essential oil in 200 mL distilled water, and the device was hidden from participants to avoid visual cueing. Pilot ratings indicated high detectability ($M=6.00/7$, $SD=0.93$), moderate-to-high intensity ($M=5.13$, $SD=1.55$), and moderate comfort ($M=4.63$, $SD=1.85$).

The pilot was used as an engineering calibration rather than a statistical pretest of the formal hypotheses. Its purpose was to avoid two obvious failure modes: a cue so weak that participants could not plausibly notice it and a cue so intense or persistent that discomfort dominated the takeover experience. Candidate preference and acceptance were therefore considered together with detectability, timing, and residual-odor comments. Reported odor detection occurred approximately 3 s after manual spray operation in the tested arrangement. This observation motivated the S3 $-3$ s operator-cue offset; it was not treated as a precise measurement of physical transport or perceptual onset.

Table~\ref{tab:pilot-design-decisions} summarizes how the pilot informed the formal apparatus. The full pilot questionnaire appears in Appendix~\ref{app:pilot-questionnaire}.

\begin{table*}[t]
\centering
\caption{Pilot calibration evidence and resulting design decisions ($N=8$). The pilot informed apparatus calibration and usability choices rather than hypothesis testing.}
\label{tab:pilot-design-decisions}

\small
\renewcommand{\arraystretch}{1.30}
\setlength{\tabcolsep}{4pt}

\begin{tabular}{
p{0.22\textwidth}
p{0.28\textwidth}
p{0.43\textwidth}
}
\hline
\textbf{Calibration evidence} &
\textbf{Pilot observation} &
\textbf{Formal-study decision} \\
\hline

Detectability &
$M=6.00$, $SD=0.93$ (1--7). &
Retain peppermint as a detectable auxiliary cue; auditory TOR remains the explicit command. \\

Intensity and comfort &
Intensity $M=5.13$, $SD=1.55$; comfort $M=4.63$, $SD=1.85$. &
Use the selected mixture without increasing intensity further. \\

Suitability / acceptance &
Suitability $M=5.25$, $SD=1.04$; willingness to accept $M=5.50$, $SD=0.93$. &
Treat scent as optional support rather than a stand-alone safety-critical signal. \\

Candidate preference &
Candidate D preferred by 6/8 pilot participants. &
Select Candidate D for the formal study. \\

Timing / carryover &
Reported detection followed manual operation by approximately 3 s; comments also raised lingering-scent concerns. &
Place the S3 operator cue 3.0 s before auditory TOR and require at least 300 s washout after odor-assisted blocks. \\

\hline
\end{tabular}
\end{table*}

\subsubsection{Temporal Alignment Framework}
Following the four-stage olfactory event model defined in Sec.~\ref{sec:timing-coordination}, the apparatus directly logged the software release cue and experimental events, while physical odor arrival and perceptual detection remained uninstrumented. Thus, S2 denotes alignment at the TOR software-control point, not perceptual simultaneity.

\begin{figure}[h]
  \centering
 \includegraphics[width=\linewidth]{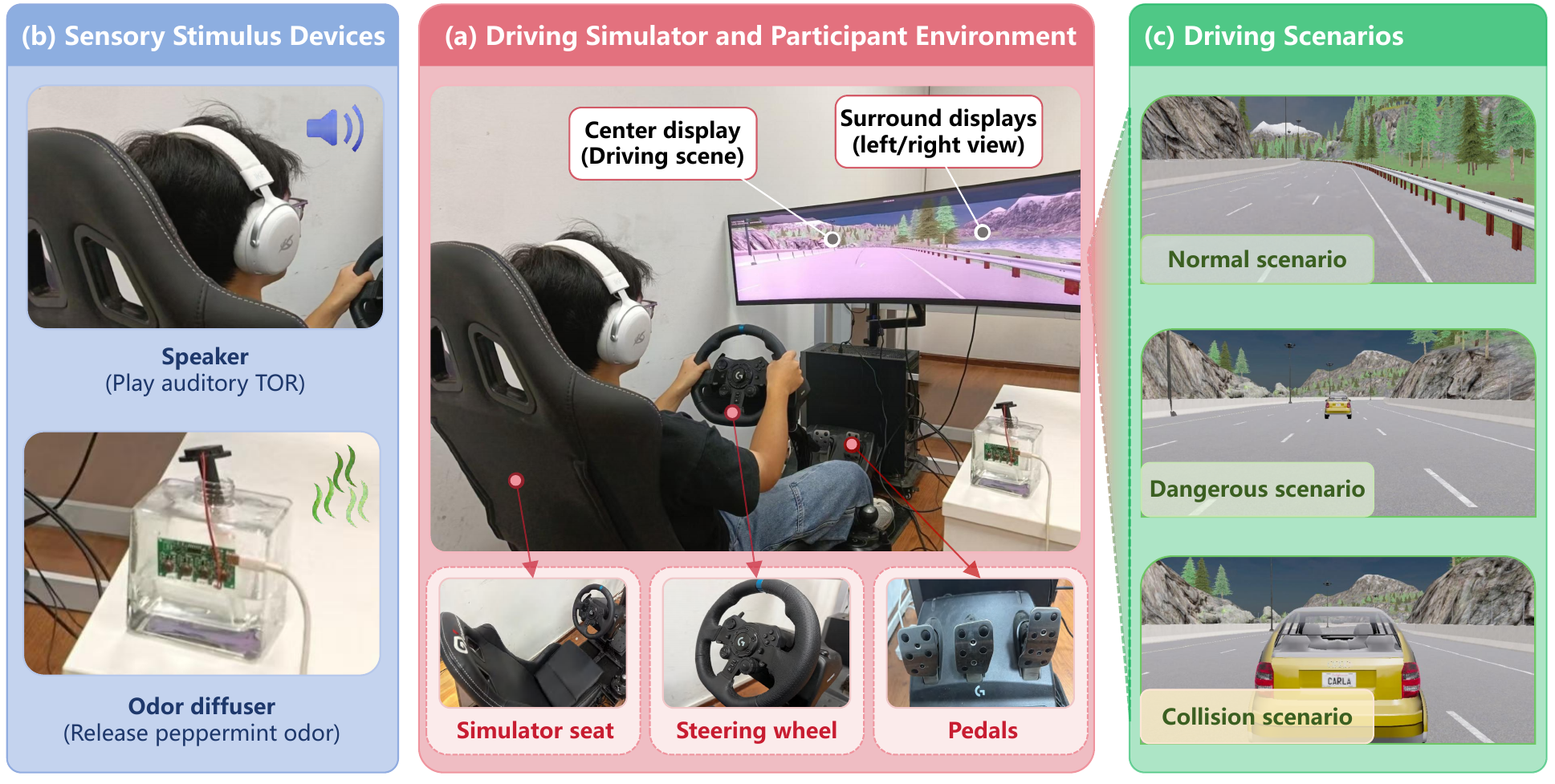}
  \caption{Experimental setup and stimulus-delivery concept. The auditory TOR is software controlled and constant across conditions. The peppermint device is manually operated by the experimenter in response to the Operator Panel, so software release-cue time is distinguished from physical and perceptual odor onset.}
  \Description{Experimental setup and stimulus delivery.}
  \label{fig:setup}
\end{figure}

For all conditions, Takeover Time uses the auditory TOR timestamp as the common origin. The earlier operator scent-release cue in S3 therefore does not move analysis time zero; no second scent release occurred at TOR, and the Operator Panel remained hidden from participants. Fig.~\ref{fig:setup} shows the apparatus and timing boundary.

\subsection{Experimental Integration}
The controller implemented the study as a state sequence and logged continuous telemetry, discrete experimental events, questionnaire responses, session metadata, and derived outcomes on a shared simulation clock. Before the formal experiment, checks verified CARLA, wheel/pedal connectivity, audio, Operator Panel state, scent-device readiness, and log creation. During each block, the experimenter monitored system health without revealing TOR timing or providing conversational cues.

Detailed block timing, KSS administration, washout, and post-block procedures are reported in Sec.~\ref{sec:procedure} and Fig.~\ref{fig:procedure}. The system-level contribution of this integration is traceability: TOR events, participant input, vehicle response, and logged delivery anomalies can be reconstructed on a common timeline.

The logging stack separates continuous and discrete evidence. Telemetry records vehicle state and control input at each synchronous tick, while event logs record transitions such as KSS presentation, odor-release cueing, TOR, and takeover detection. Questionnaire and session metadata provide condition labels and subjective outcomes. This structure supports retrospective checks for logging or protocol anomalies, although it cannot recover trials absent from the frozen dataset.

\section{Experimental Design}
\subsection{Participants and Screening}
The formal study recruited 18 participants through an open call posted on the university recruitment platform. Participants received RMB~50 compensation. The sample comprised 13 men and 5 women, with a mean age of 23.4 years. The protocol was designed as a single-factor within-participant study in which each participant would complete all three reminder conditions; technical interruptions during data collection resulted in incomplete condition coverage for the realized dataset.

Before training, participants completed health and safety screening for nasal congestion or cold symptoms, known allergies, asthma or respiratory irritation risk, severe motion-sickness tendency, alcohol use, medication affecting alertness, and strong fragrance exposure. Sleep, caffeine/smoking, driving experience, and automated-driving familiarity were recorded where available. Participants provided informed consent and could pause or withdraw at any time.

\subsection{Apparatus and Driving Scenario}
The study used CARLA 0.9.15 on Town04 in synchronous mode with a 0.05-s step (nominally 20 Hz) and a target automated-driving speed of 90 km/h. The analyzed build had no background NPC vehicles or pedestrians. Participants used a Logitech G923 wheel/pedal set; the driving view was 3840$\times$1080, with a 1920$\times$540 CARLA camera and 100-degree field of view.

All bundles used Town04 road 45, lane $-4$, and a stationary forward vehicle. Shared route geometry and hazard class reduce variability, but condition and scenario identifier were fixed within bundles rather than independently crossed. The frozen export also lacks a condition-independent configuration table that would verify numerical equality of every hazard parameter post hoc, so observed differences cannot be attributed solely to reminder timing.

The simplified scene prioritized repeatability over realism. Its high collision proportions (Sec.~\ref{sec:collision}) indicate a demanding stress case rather than a representative sample of routine Level 3 handovers. The lack of traffic also affects interpretation of lateral movement, which can reflect deliberate obstacle avoidance. Generalization therefore requires replication across broader hazard horizons, traffic densities, and maneuver options.

\subsection{Conditions, Counterbalancing, and Procedure}
\label{sec:conditions}
\subsubsection{Conditions}
At the reminder-programming level, the three arrangements varied the presence and timing of olfactory support relative to a fixed auditory TOR. The formal protocol also bundled each reminder arrangement with a predefined scenario identifier, so the realized experiment was not a timing-only manipulation.
\begin{itemize}
    \item \textbf{Audio-only (S1):} auditory TOR with no scent-release cue.
    \item \textbf{TOR-Control-Point Aligned (S2):} operator scent-release cue at the auditory TOR software-control point.
    \item \textbf{$-3$ s Operator-Cue Lead (S3):} operator scent-release cue 3.0 s before the auditory TOR, with no second release at TOR.
\end{itemize}
The auditory TOR remained the common analysis time origin. Although the protocol intended three blocks per participant, the realized dataset is not a complete repeated-measures sample.

\subsubsection{Counterbalancing}
Six order groups covered all $3!=6$ permutations of the condition--scenario bundles, distributing conditions across early, middle, and late positions. Because technical interruptions left 21 planned slots without analyzable trial records, the realized dataset no longer preserves the intended balance, so counterbalancing is reported as protocol design rather than evidence for a causal within-participant comparison.

Odor-assisted blocks used a minimum 300-s ventilation/washout, with the next block beginning only after residual odor was judged sufficiently low and no relevant discomfort was reported. Practice included manual driving, automated driving, and one Audio-only TOR, but no formal odor cue or hazard scenario.

\subsubsection{Experimental Procedure}
\label{sec:procedure}
\begin{figure}[h]
  \centering
 \includegraphics[width=\linewidth]{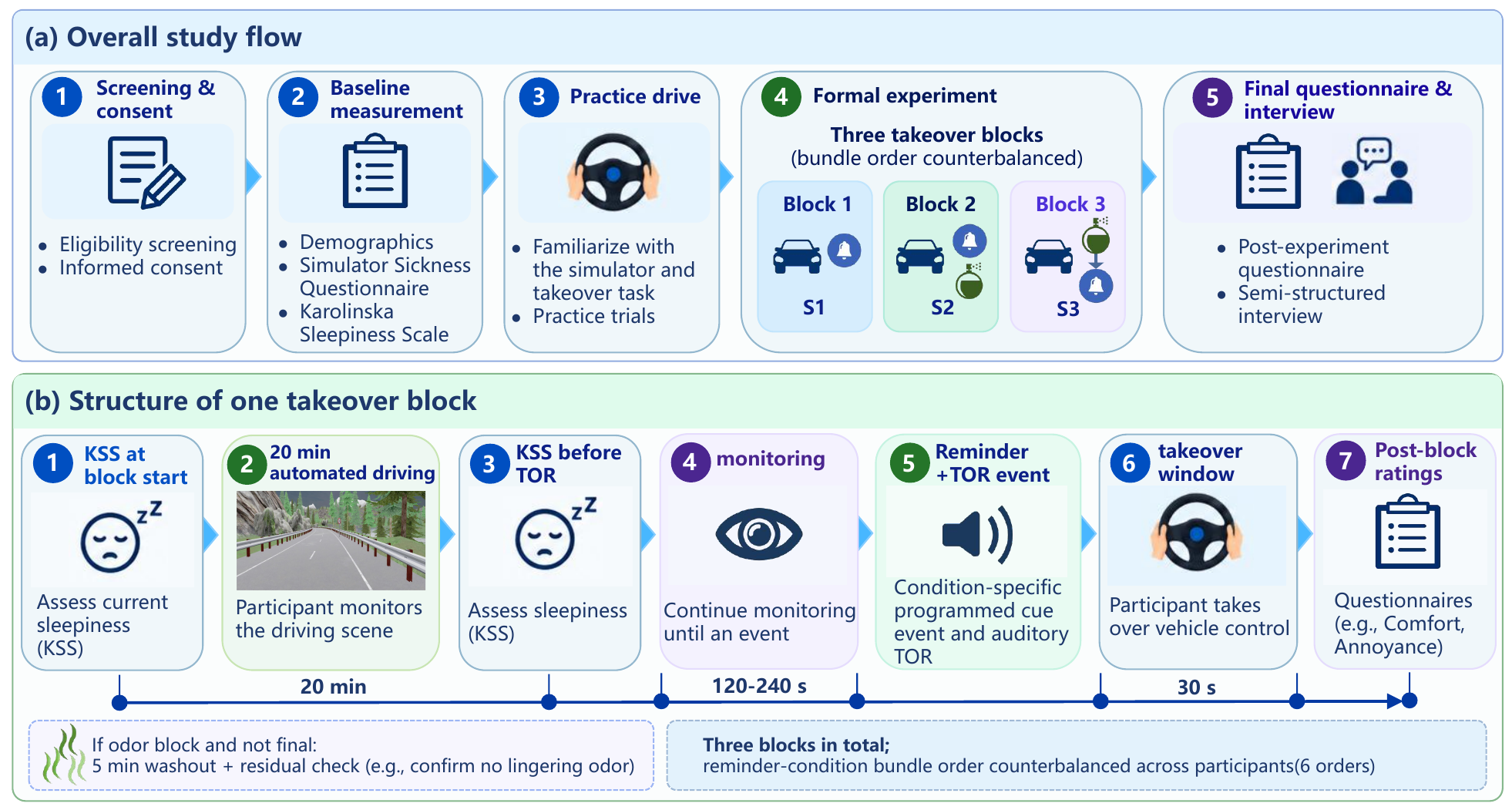}
  \caption{Experimental procedure. Each block comprised KSS at block start, 20 min of monotonous automated driving, KSS before TOR, a randomized 120--240 s interval, the condition-specific reminder sequence, takeover, and a post-block questionnaire. Odor-assisted blocks were followed by at least 300 s of washout.}
  \Description{Experimental procedure.}
  \label{fig:procedure}
\end{figure}

Each session began with screening, informed consent, baseline KSS, simulator-discomfort assessment, seat/control adjustment, and 3--5 min of practice. Fig.~\ref{fig:procedure} summarizes the block sequence.

Each formal block comprised KSS-BlockStart; 20 min of monotonous automated driving with basic road monitoring and no additional non-driving task; KSS-PreTOR; a randomized 120--240-s automated-driving interval; the condition-specific reminder and auditory TOR; a 30-s takeover observation window; and the post-block questionnaire. The randomized interval reduced TOR predictability. The experimenter did not reveal TOR timing or provide steering/braking advice after TOR. Odor-assisted blocks were followed by washout; after the formal blocks, participants completed a final questionnaire, simulator-discomfort check, and short semi-structured interview.

\subsection{Measures}
We evaluated behavior, safety, vehicle control, state, and experience. KSS (1 = extremely alert, 9 = extremely sleepy) was administered at baseline, block start, and pre-TOR \cite{aakerstedt1990subjective,kaida2006validation}; simulator discomfort was checked pre/post using the study's SSQ procedure \cite{kennedy1993simulator}.

Takeover Time was measured from auditory TOR onset to the first sustained valid manual input: steering change $\geq2^\circ$ from the local pre-TOR baseline or normalized brake input $\geq0.10$, sustained for $\geq100$ ms \cite{zhang2019determinants,zeeb2016take,louw2017coming}. Minimum TTC was the smallest valid pre-collision TTC after TOR \cite{hayward1972near}; Collision was binary. Maximum deceleration used the most negative accepted local speed slope over trailing 0.50-s windows, and maximum absolute lane-center offset measured displacement from the nearest current CARLA lane center rather than SDLP.

After each available block, participants rated Comfort, Annoyance, Acceptance, and Perceived Effectiveness on 1--7 scales. The post-study interview compared reminder experiences, peppermint, perceived effectiveness, long-term acceptability, and suggested improvements. Table~\ref{tab:measure-framework} summarizes operational definitions and evidence boundaries; Appendix~\ref{app:interview-guide} gives the interview guide.

\begin{table*}[t]
\centering
\caption{Outcome measures and interpretation boundaries. Sample sizes vary by metric because the frozen dataset is incomplete.}
\label{tab:measure-framework}

\small
\renewcommand{\arraystretch}{1.28}
\setlength{\tabcolsep}{4pt}

\begin{tabular}{
p{0.18\textwidth}
p{0.34\textwidth}
p{0.41\textwidth}
}
\hline
\textbf{Measure} &
\textbf{Operationalization} &
\textbf{Interpretation boundary} \\
\hline

KSS &
1--9 sleepiness rating at baseline, block start, and pre-TOR. &
State context; not a direct physiological or attentional measure. \\

Takeover Time &
Auditory TOR to first valid input: steering change $\geq2^{\circ}$ or brake $\geq0.10$, sustained $\geq100$ ms. &
Onset of manual intervention; not full situation-awareness recovery. \\

Minimum TTC / Collision &
Minimum valid pre-collision TTC; binary post-TOR collision outcome. &
Safety margin and event outcome are complementary, not interchangeable. \\

Maximum deceleration &
Most negative accepted local speed slope over trailing 0.50-s windows. &
Longitudinal intervention intensity; stronger braking is not automatically better or worse. \\

Lane-center offset &
Maximum absolute displacement from the nearest current CARLA lane center. &
Lateral movement; not SDLP or a pure lane-keeping-quality score. \\

Subjective ratings &
1--7 Comfort, Annoyance, Acceptance, and Perceived Effectiveness. &
Perceived experience; does not substitute for objective safety outcomes. \\

Interview &
Short post-study semi-structured interview. &
Contextual interpretation only; does not establish unmeasured mechanisms or theme prevalence. \\

\hline
\end{tabular}
\end{table*}

\subsection{Data Quality and Analysis}
The planned design yields 54 condition slots (18$\times$3); 33 trial rows are available (61.1\%), leaving 21 planned slots without analyzable records: S1 is available in 12 sessions, S2 in 13, and S3 in 8. Study-administration records attribute the unavailable slots to equipment-related technical interruptions during data collection rather than performance-based trial exclusion or participant withdrawal. Detailed fault categories were not retained for every affected block, so we cannot establish that interruption probability was independent of condition. The session-level map is reported in Appendix~\ref{app:data-audit}, Table~\ref{tab:condition-slot-flow}.

This structure does not support a conventional complete-case repeated-measures ANOVA, and stronger model-based inference would depend on assumptions the records cannot verify. We therefore report exact denominators, descriptive statistics, and 10,000-resample percentile bootstrap 95\% confidence intervals. These intervals are descriptive uncertainty summaries conditional on the observed trials: they do not correct condition imbalance or the condition--scenario confound. For cells with $n<10$, their nominal 95\% population coverage should not be assumed; in particular, the Cue-Lead (S3) Takeover Time interval ($n=6$) is not used for inferential interpretation.

Interview data were examined through a lightweight structured reading organized around domains in the interview guide (preparation, directness, combined salience, comfort/annoyance, scent experience, and long-term use). The analysis record does not document an independent second coder or inter-rater statistic, so qualitative material is used only to contextualize quantitative patterns; we do not report theme prevalence or causal attentional/physiological mechanisms.

\subsection{Ethics and Safety}
The study has obtained ethical approval from the institution’s ethics committee$^1$.
\footnotetext[1]{Due to the requirement of anonymity for manuscript submission, the name of the institution is temporarily not disclosed.}
All participants provided informed consent and could pause or withdraw at any time. Screening addressed olfactory/respiratory sensitivity and motion-sickness risk; odor presentation was to stop and the environment be ventilated if discomfort occurred. Simulator discomfort was monitored throughout the visit. Data were stored under anonymous identifiers and backed up in redundant raw records.

\section{Results}
\subsection{Final Dataset, Missingness, and Uncertainty}
The frozen dataset contains 33 available TOR trials from 18 participants: 12 Audio-only (S1), 13 Aligned (S2), and 8 Cue-Lead (S3). Twenty-one of 54 planned condition slots did not yield analyzable trial records because of technical interruptions during data collection. Only two participants have all three conditions. Takeover Time is available for 26 trials (S1 $n=9$, S2 $n=11$, S3 $n=6$); TTC, Collision, maximum deceleration, and lane-center offset for all 33; and the four primary subjective ratings for 30 condition observations.

The interruptions were not performance-based exclusions, but detailed per-block fault categories were not retained and availability was uneven across conditions. We therefore cannot establish that interruption probability was independent of condition. Appendix~\ref{app:data-audit} reports the session-level pattern, and bootstrap intervals in Appendix~\ref{app:descriptive-statistics} quantify uncertainty among observed trials only. S3's smaller available sample is therefore interpreted especially cautiously.

\subsection{RQ1: Behavioral and Safety Outcomes}
\subsubsection{Takeover Time}

Audio-only had a mean Takeover Time of 3.78 s ($SD=2.68$, $n=9$), Aligned (S2) 3.18 s ($SD=1.92$, $n=11$), and Cue-Lead (S3) 2.12 s ($SD=0.96$, $n=6$). The corresponding medians were 2.76, 2.86, and 2.23 s, and bootstrap 95\% intervals were [2.19, 5.48], [2.11, 4.25], and [1.42, 2.80] s. S3 had the shortest observed response time, but the small $n$ and overlapping uncertainty preclude a confirmatory claim.

Takeover Time uses the auditory TOR as the common origin, so S3's $-3$ s operator cue lies outside the response interval. This ordering should not be attributed to an early experienced odor: perceptual onset was unmeasured, retrospective S3 detection was sparse, condition was bundled with scenario, and the retained S3 sample was small. Preparation is therefore only a candidate explanation for replication.

The narrower S3 interval does not imply greater population precision. With only six values, percentile-bootstrap coverage cannot be assumed nominal and does not capture uncertainty from unavailable S3 slots; we therefore use it descriptively rather than inferentially.

\subsubsection{Minimum TTC and Collision}
\label{sec:collision}
Minimum TTC averaged 2.36 s ($SD=2.47$, $n=12$) for Audio-only, 1.88 s ($SD=1.99$, $n=13$) for Aligned (S2), and 2.98 s ($SD=2.11$, $n=8$) for Cue-Lead (S3); bootstrap 95\% intervals were [1.10, 3.69], [0.89, 2.99], and [1.58, 4.25] s. Cue-Lead (S3) thus shows the largest observed TTC, while Aligned (S2) shows the smallest.

Collision proportions were high in every condition: 6/12 (50.0\%) in Audio-only, 7/13 (53.8\%) in Aligned (S2), and 3/8 (37.5\%) in Cue-Lead (S3). Bootstrap intervals for the observed proportions were correspondingly wide: [25.0\%, 75.0\%], [30.8\%, 76.9\%], and [12.5\%, 75.0\%]. The lower Cue-Lead (S3) proportion and larger TTC point in the same descriptive direction, but the high absolute collision rate and uncertainty are at least as important as the ordering.

In other words, the dataset does not support a claim that any reminder configuration made the scenario safe. The collision outcome is better used here to characterize the severity of the test context and to motivate a replication with multiple hazard horizons than to rank the three bundles. They indicate that this experiment used a demanding hazard context and constrain how broadly the pattern can be generalized.

\subsubsection{Vehicle Control after TOR}
Maximum deceleration averaged $-5.53$ m/s$^2$ in Audio-only, $-6.48$ m/s$^2$ in Aligned (S2), and $-4.34$ m/s$^2$ in Cue-Lead (S3). Maximum absolute lane-center offset averaged 0.57, 1.58, and 1.73 m, respectively. Thus, Cue-Lead (S3) combined milder average braking with the largest lateral displacement, whereas Audio-only showed the smallest lateral displacement. Because the lane metric is relative to the nearest current lane center, it can reflect an intentional evasive movement as well as less stable control; it is not a simple lane-keeping score.

Taken together, RQ1 does not yield a single dominant arrangement. S3 Cue-Lead has the most favorable observed ordering on Takeover Time, TTC, Collision, and mean deceleration, but it also has the largest lateral offset and the smallest retained sample. S2 Aligned combines stronger braking and smaller TTC with intermediate response time, whereas S1 shows slower response but less lateral displacement. The value of this profile is therefore demonstrative rather than comparative: different outcome families can point in different directions even before the unmeasured perceptual timing and the condition--scenario bundle are considered. Fig.~\ref{fig:behavior} visualizes these descriptive patterns; exact values and bootstrap intervals are listed in Appendix~\ref{app:descriptive-statistics}.

\begin{figure}[h]
  \centering
 \includegraphics[width=\linewidth]{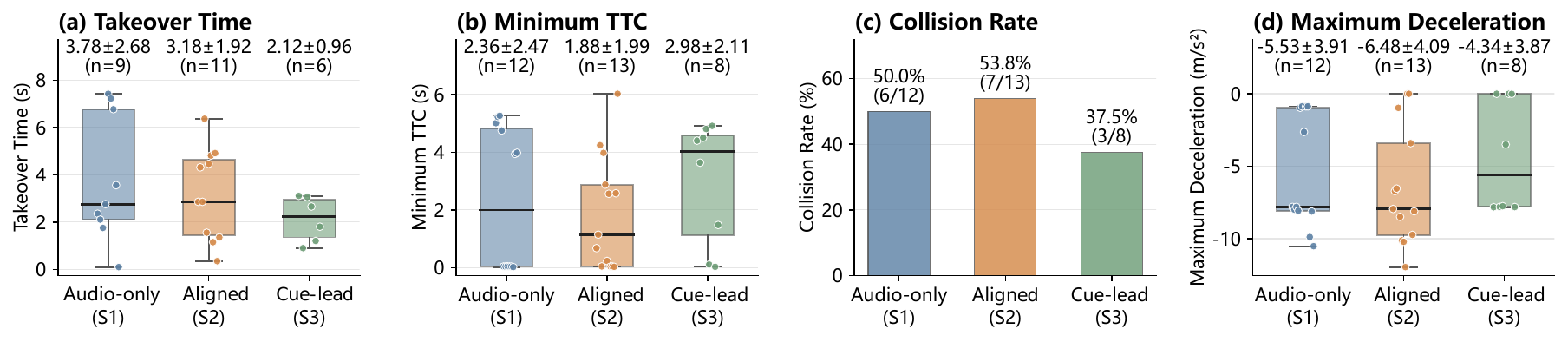}
  \caption{Behavioral and safety outcomes across the three programmed reminder arrangements. The four panels summarize Takeover Time, Minimum TTC, Collision, and maximum deceleration. Maximum absolute lane-center offset is reported in the text and Appendix Table~\ref{tab:supp-behavior}.}
  \Description{Four-panel comparison of Takeover Time, Minimum TTC, Collision, and maximum deceleration across S1, S2, and S3.}
  \label{fig:behavior}
\end{figure}

\subsection{RQ2: Experience, Detection, and Interpretation Boundaries}
\subsubsection{Comfort, Annoyance, Acceptance, and Perceived Effectiveness}
Comfort averaged 3.33 for Audio-only, 4.27 for Aligned (S2), and 4.43 for Cue-Lead (S3). Annoyance averaged 4.08, 3.09, and 3.86, respectively. The odor-assisted arrangements were therefore more comfortable on average than Audio-only, while Aligned (S2) was least annoying among the observed responses.

Acceptance was similar across conditions (4.67 Audio-only, 4.64 Aligned (S2), 4.57 Cue-Lead (S3)). Perceived Effectiveness showed a different ordering: 4.08, 5.09, and 4.57, with Aligned (S2) highest. Bootstrap intervals for these subjective means overlap substantially (Appendix Table~\ref{tab:supp-subjective}), reinforcing that the main point is divergence across outcome types rather than a reliable rank order. In particular, Aligned (S2) was perceived as most effective even though its observed TTC and collision pattern was not the most favorable.

This discrepancy is informative for HCI evaluation because perceived effectiveness may reflect salience, clarity, and confidence rather than the simulator-recorded margin. Conversely, an earlier cue may shape how participants approach the TOR without producing the strongest subjective impression at command onset.

\subsubsection{Qualitative Context and Long-Term Use}
The qualitative material provides context for this divergence but is not treated as independent evidence of perceptual order or as a prevalence-based thematic result. The post-study interview used condition-specific prompts that referred to the intended/programmed temporal relation (Appendix~\ref{app:interview-guide}); because physical odor arrival and perceptual onset were not measured, this wording may have shaped participants' retrospective reconstruction of the sequence. Available comments are therefore used only to generate candidate interpretations---for example, directness around S1, concentrated multimodal salience around S2, or possible preparation around S3---rather than to establish that these functions occurred. We do not report theme counts because transcript/session coverage and a second-coder procedure are not documented well enough to support frequency claims.

Comments about long-term use shifted attention from immediate salience to scent preference, residual odor, repeated exposure, and the ability to personalize or disable the olfactory channel. These concerns matter because scent changes the cabin environment beyond the instant of the TOR. The current short-term study cannot establish whether these reactions would persist, habituate, or reverse over repeated real-world use.

An ancillary negative-control check is also informative: in Audio-only, no respondent selected yes to odor detection (0 Yes, 11 No, 1 Uncertain; Appendix Table~\ref{tab:odor-audit}). This pattern does not rule out expectancy effects, but it argues against systematic false-positive reports of smelling the cue when no scent was released.

\begin{figure}[h]
  \centering
 \includegraphics[width=\linewidth]{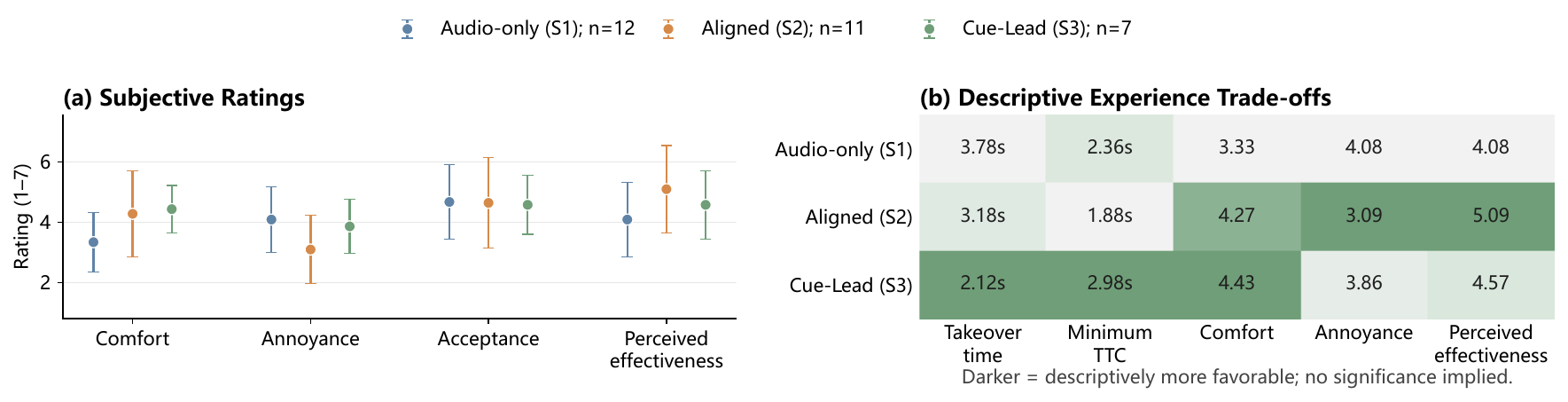}
  \caption{Driver experience and interpretation boundaries. Panels summarize Comfort, Annoyance, Acceptance, and Perceived Effectiveness. Qualitative annotations are shown only as prompt-sensitive candidate interpretations concerning directness, combined salience, possible preparation, odor preference, and long-term use; they are not treated as measured mechanisms or prevalence-based themes.}
  \Description{Comparison of subjective ratings and qualitative context across the three programmed reminder arrangements.}
  \label{fig:experience}
\end{figure}

As shown in Fig.~\ref{fig:experience}, subjective experience does not reproduce the ordering of the safety-related measures. The result supports RQ2 as a trade-off question: a combined cue may feel more effective or less annoying, while an earlier auxiliary cue may be associated with a different behavioral pattern. Neither observation establishes a universally preferred or safer design.

\section{Discussion}
\label{sec:discussion}

The study's strongest contribution is not evidence that one programmed arrangement is superior. Rather, the formal study demonstrates why olfactory--auditory timing requires explicit evidence accounting. The behavioral and experience data provide exploratory comparison, while the detection audit exposes a more basic construct-validity issue: a reproducible software/operator offset cannot be assumed to be the interval experienced by the driver.

\subsection{RQ1: Programmed Arrangements Produce Descriptive, Not Causal, Outcome Profiles}
RQ1 asked what behavioral, safety, and vehicle-control patterns were observed across the three programmed arrangements. Fig.~\ref{fig:behavior} shows that the retained trials did not yield a common ordering across response, safety margin, collision, and control behavior. S3 had the shortest observed Takeover Time, largest TTC, and lowest collision proportion, but also the largest lateral displacement and the smallest sample. S2 had intermediate response time, the smallest TTC, and strongest average deceleration. S1 showed slower response but the smallest lateral displacement.

These patterns are useful as a demonstration dataset because they show which outcomes can diverge when a slow auxiliary modality is repositioned in the control sequence. They are not evidence that an early perceived odor caused the S3 pattern. The formal study cannot establish that claim for three independent reasons: the perceptual lead was not measured, condition was bundled with scenario, and the intended repeated-measures structure was substantially disrupted by technical interruptions. RQ1 therefore contributes candidate observations for replication rather than a ranking of timing strategies.

\subsection{RQ2: Participant Experience Exposes the Gap between Control Timing and Experienced Cueing}
RQ2 adds an important layer because the subjective data and manipulation check do not map cleanly onto the programmed labels. S2 received the highest Perceived Effectiveness and lowest Annoyance ratings, while S3 was highest on Comfort. More importantly for the methodological argument, conscious odor detection was reported much more often in S2 than in S3 even though S3 was the arrangement with the earlier operator cue. Because detection was retrospective and untimed, this does not reveal when odor arrived. It does, however, show that a $-3$ s programmed operator-cue lead should not be described as a reliable perceptual pre-cue.

The interviews are weaker evidence for perceptual ordering than the condition-specific wording might initially suggest. The guide referred to the intended/programmed relation using phrases such as the sound and odor being initiated together and the odor appearing before the sound (Appendix~\ref{app:interview-guide}), even though the experiment did not measure physical or perceptual onset. This creates a plausible interviewer-framing or demand-characteristic pathway through which participants could retrospectively organize their experience around the intended sequence. We therefore treat the interview material only as a source of candidate hypotheses. Some responses were compatible with possible preparation before the explicit TOR, while others were compatible with concentrated multimodal salience around the TOR; neither pattern establishes a cognitive mechanism or verifies the temporal order participants actually experienced.

\subsection{Methodological Synthesis: A Temporal Evidence-Accounting Workflow}
\label{sec:evidence-accounting-workflow}
Prior work has already established olfactory takeover support as a legitimate design space; Tang et al., for example, evaluated peppermint alongside auditory and tactile takeover requests \cite{tang2021olfactory}. Likewise, two-stage warning research has already shown the value of separating a preparatory stage from a later command \cite{zhang2025effects,tan2026there,wu2025request}. Our incremental contribution is therefore neither olfaction for takeover nor two-stage warning by itself. It is a reusable procedure for specifying what a temporal relation refers to, what evidence supports it, and which interpretations remain justified when a modality has a non-negligible delivery chain.

Based on the four-stage model in Sec.~\ref{sec:timing-coordination}, we operationalize \textit{temporal evidence accounting} as a five-step workflow. The workflow is intended as a reporting-and-inference procedure rather than a validated sensing standard:
\begin{enumerate}
    \item \textbf{Enumerate modality-specific event stages.} Identify the relevant control, actuation, physical delivery/arrival, perceptual detection, and response events for each channel rather than collapsing them into a single onset time.
    \item \textbf{Assign an evidence status to each stage.} Mark each event as controlled, timestamped/logged, directly measured, retrospectively reported, or unobserved. An event should not be promoted to a stronger evidence status through interpretation alone.
    \item \textbf{Name conditions by the relation that is actually controlled or established.} Programmed or logged relations can support control-level labels; physical or perceptual labels require corresponding physical or perceptual evidence.
    \item \textbf{Define a common analysis reference and preserve trial provenance.} Specify the event used as the temporal origin for outcome measures and retain the planned assignment, cue-delivery status, sensing status, interruption/exclusion status, and archival outcome for each trial.
    \item \textbf{Bound each claim to the evidence that supports the relevant relation.} Programmed offsets support claims about programmed timing; retrospective detection supports whether a cue was reported as noticed, not when it was first perceived; behavioral outcomes support observed response patterns, not an unmeasured cognitive mechanism.
\end{enumerate}

In the present study, the auditory TOR provides the common analysis reference, while S2 and S3 are named from the software/operator cue relation that the apparatus controls. Manual-actuation latency, physical odor arrival, and perceptual onset are not directly observed; the post-block Detection item adds retrospective evidence about reported noticing but not onset time. The detection audit therefore illustrates the workflow's practical value: calling S3 ``odor before audio'' would convert a programmed $-3$ s offset into an unsupported perceptual claim, while calling S2 ``simultaneous'' would similarly exceed the available evidence. Table~\ref{tab:tradeoff-synthesis} provides a compact instance of the resulting evidence ledger by linking each programmed relation to observed evidence and an explicit interpretation boundary.

The workflow does not repair missing measurements or make an incomplete experiment causal. Its purpose is to make the inferential consequences of those gaps visible and reproducible. Although motivated here by olfactory--auditory TORs, the procedure applies more generally to multimodal systems in which a controlled trigger can be separated from physical or perceptual onset by non-negligible or variable latency. The CARLA instrument is one implementation of this logic: it provides traceability across the auditory reference, programmed scent cue, participant input, vehicle response, manipulation-check data, and known unmeasured stages. A stronger implementation should add automated scent actuation, concentration/arrival sensing near the driver, and, where feasible, a perceptual-onset measure.

\subsection{Design Considerations and Testable Hypotheses}
The following considerations are deliberately framed as requirements and hypotheses for future work rather than design rules.

\subsubsection{Name Conditions by Controlled Events Until Perceptual Onset Is Measured}
\noindent\textbf{Design Consideration 1:} Multimodal studies should label timing conditions according to the event they actually control. Terms such as simultaneous or odor-before-audio should be reserved for studies that measure or tightly control the relevant physical/perceptual onsets. In the present study, \textit{TOR-Control-Point Aligned} and \textit{$-3$ s Operator-Cue Lead} describe the evidence more accurately than perceptual labels.

\subsubsection{Test Candidate Functions with Timing, Urgency, and Driver State Crossed}
\noindent\textbf{Design Consideration 2:} Possible preparation and immediate multimodal salience should be treated as candidate functions to be tested, not mechanisms inferred from the present interviews. A future factorial study should cross verified perceptual timing with takeover urgency and driver state, while using neutral interview wording and, where feasible, online measures of attention or preparation. This would test whether a reliably perceived early cue supports preparation when time is available and whether concentrated multimodal stimulation is valued or behaviorally useful when urgency is high.

\subsubsection{Treat Trial Provenance as Part of Timing Methodology}
\noindent\textbf{Design Consideration 3:} Timing studies need provenance for every planned trial, not only successfully recorded trials. When whole blocks are absent, a nominal within-participant design can become a set of partially overlapping small groups. Future systems should log assignment, cue delivery status, physical-sensing status, interruption/fault category, exclusion, and archival status for each planned block. This is necessary to determine whether an apparent timing pattern could reflect selective availability rather than the manipulation itself.

Table~\ref{tab:tradeoff-synthesis} summarizes the distinction between programmed timing, what was observed, and what remains only a hypothesis.

\begin{table*}[t]
\centering
\caption{Temporal evidence accounting across the three programmed reminder arrangements. Counts are descriptive; availability varies by metric.}
\label{tab:tradeoff-synthesis}

\small
\renewcommand{\arraystretch}{1.85}
\setlength{\tabcolsep}{4pt}

\begin{tabular}{
p{0.13\textwidth}
p{0.22\textwidth}
p{0.27\textwidth}
p{0.31\textwidth}
}
\hline
\textbf{Arrangement} &
\textbf{Programmed relation} &
\textbf{Observed evidence} &
\textbf{Interpretation boundary} \\
\hline

\textbf{S1 Audio-only} &
Auditory TOR; no odor-release cue. &
Odor Detection: 0 Yes, 11 No, 1 Uncertain. &
Baseline explicit action trigger; no olfactory timing claim. \\

\textbf{S2 Aligned} &
Operator scent-release cue at the auditory TOR software-control point. &
Detection: 8 Yes, 2 No, 1 Uncertain, 2 Missing. Highest observed Perceived Effectiveness and lowest Annoyance. &
Supports control-point alignment, but not perceptual simultaneity. Salience-related interpretations remain prompt-sensitive hypotheses. \\

\textbf{S3 Cue-Lead} &
Operator scent-release cue 3.0 s before the auditory TOR. &
Detection: 2 Yes, 5 No, 1 Missing. Favorable descriptive ordering on several behavioral and safety measures. &
Supports a programmed cue lead, but not an early perceived odor or causal timing effect. Preparation-related interpretations remain hypotheses for replication. \\

\hline
\end{tabular}
\end{table*}

\subsection{Limitations and Replication Requirements}
First, 21 of 54 planned condition slots did not yield analyzable trials because of technical interruptions. Only two participants have complete retained observations for all three conditions, and S3 has the most unavailable slots. Detailed fault categories were not retained for every affected block, so condition-related availability bias cannot be excluded.

Second, condition and scenario were bundled rather than fully crossed. Although the bundles shared a broad road and hazard class, bundle-specific characteristics may contribute to the observed differences. A replication should independently cross reminder timing and scenario, or hold the scenario identical while varying timing.

Third, the study controls software/operator timing but does not measure physical odor arrival or perceptual onset. The post-block Detection item is retrospective and categorical, not a timing measure. This is precisely the boundary the methodological framing is intended to expose, but it also prevents the current experiment from validating a perceptual timing effect.

Fourth, qualitative material was examined through a lightweight structured interpretive reading without a documented independent second coder or inter-rater statistic. The candidate concepts of anticipatory buffering and immediate multimodal salience should therefore be read as provisional interpretations rather than prevalence-based themes or validated mechanisms.

Finally, collision proportions were high (37.5--53.8\%), indicating a demanding stationary-hazard context. The study also used one peppermint mixture, one programmed lead, a simplified simulator scene, short-term exposure, and a demographically narrow university-platform sample. Broader claims require larger, fully crossed replications with verified scent delivery, multiple hazard horizons, varied driver states, and complete trial provenance.

\section{Conclusion}
This work reframes olfactory--auditory takeover reminders as a temporal evidence problem rather than a comparison of programmed timing strategies. We contribute an event-level representation and a five-step temporal evidence-accounting workflow that distinguish controlled cue timing from manual actuation, physical odor arrival, and perceptual onset. Our exploratory dataset shows why this distinction matters: a programmed $-3$ s operator-cue lead does not establish an experienced olfactory lead, and control-point alignment does not establish perceptual simultaneity. Accordingly, observed behavioral, safety, and subjective patterns are treated as hypothesis-generating rather than causal evidence. Future studies should combine fully crossed designs with instrumented scent delivery and, where feasible, direct measures of perceptual onset.

\bibliographystyle{ACM-Reference-Format}
\bibliography{ref}

@book{on2021taxonomy,
  title={Taxonomy and definitions for terms related to driving automation systems for on-road motor vehicles},
  author={On-Road Automated Driving (ORAD) Committee},
  year={2021},
  publisher={SAE international}
}

@article{parasuraman1997humans,
  title={Humans and automation: Use, misuse, disuse, abuse},
  author={Parasuraman, Raja and Riley, Victor},
  journal={Human factors},
  volume={39},
  number={2},
  pages={230--253},
  year={1997},
  publisher={SAGE Publications Sage CA: Los Angeles, CA}
}

@article{endsley1995out,
  title={The out-of-the-loop performance problem and level of control in automation},
  author={Endsley, Mica R and Kiris, Esin O},
  journal={Human factors},
  volume={37},
  number={2},
  pages={381--394},
  year={1995},
  publisher={SAGE Publications Sage CA: Los Angeles, CA}
}

@article{endsley1995toward,
  title={Toward a theory of situation awareness in dynamic systems},
  author={Endsley, Mica R},
  journal={Human factors},
  volume={37},
  number={1},
  pages={32--64},
  year={1995},
  publisher={SAGE Publications Sage CA: Los Angeles, CA}
}

@article{parasuraman2000model,
  title={A model for types and levels of human interaction with automation},
  author={Parasuraman, Raja and Sheridan, Thomas B and Wickens, Christopher D},
  journal={IEEE Transactions on systems, man, and cybernetics-Part A: Systems and Humans},
  volume={30},
  number={3},
  pages={286--297},
  year={2000},
  publisher={IEEE}
}

@article{lee2004trust,
  title={Trust in automation: Designing for appropriate reliance},
  author={Lee, John D and See, Katrina A},
  journal={Human factors},
  volume={46},
  number={1},
  pages={50--80},
  year={2004},
  publisher={SAGE Publications Sage UK: London, England}
}

@article{pan2025exploring,
  title={Exploring the effect of driver drowsiness on takeover performance during automated driving: An updated literature review},
  author={Pan, Hengyan and Logan, David B and Stephens, Amanda N and Payre, William and Wang, Yonggang and Peng, Zhipeng and Qin, Yang and Koppel, Sjaan},
  journal={Accident Analysis \& Prevention},
  volume={216},
  pages={108023},
  year={2025},
  publisher={Elsevier}
}

@article{chen2025systematic,
  title={A systematic review on test performance of the driver takeover process in automated driving},
  author={Chen, Haolin and Zhao, Xiaohua and Chen, Chen and Li, Zhenlong and Li, Haijian and Wang, Qiuhong},
  journal={Accident Analysis \& Prevention},
  volume={215},
  pages={108012},
  year={2025},
  publisher={Elsevier}
}

@article{zhang2019determinants,
  title={Determinants of take-over time from automated driving: A meta-analysis of 129 studies},
  author={Zhang, Bo and De Winter, Joost and Varotto, Silvia and Happee, Riender and Martens, Marieke},
  journal={Transportation research part F: traffic psychology and behaviour},
  volume={64},
  pages={285--307},
  year={2019},
  publisher={Elsevier}
}

@article{liang2026towards,
  title={Towards safe and comfortable vehicle control transitions: A systematic review of takeover time, time budget, and takeover outcomes},
  author={Liang, Kexin and Calvert, Simeon C and van Lint, JWC},
  journal={Human Factors},
  pages={00187208261456057},
  year={2026},
  publisher={SAGE Publications Sage CA: Los Angeles, CA}
}

@article{wu2025request,
  title={How to request drivers to prepare for takeovers during automated driving},
  author={Wu, Yanbin and Hasegawa, Kunihiro and Kihara, Ken},
  journal={Transportation Research Part F: Traffic Psychology and Behaviour},
  volume={109},
  pages={938--950},
  year={2025},
  publisher={Elsevier}
}

@article{tan2026there,
  title={When there is no rush to take over: Evaluating multi-stage takeover request designs in conditionally automated driving},
  author={Tan, Xiaomei and Zhang, Yiqi},
  journal={Transportation Research Part F: Traffic Psychology and Behaviour},
  volume={117},
  pages={103465},
  year={2026},
  publisher={Elsevier}
}

@article{zeeb2015determines,
  title={What determines the take-over time? An integrated model approach of driver take-over after automated driving},
  author={Zeeb, Kathrin and Buchner, Axel and Schrauf, Michael},
  journal={Accident analysis \& prevention},
  volume={78},
  pages={212--221},
  year={2015},
  publisher={Elsevier}
}

@article{zeeb2016take,
  title={Is take-over time all that matters? The impact of visual-cognitive load on driver take-over quality after conditionally automated driving},
  author={Zeeb, Kathrin and Buchner, Axel and Schrauf, Michael},
  journal={Accident analysis \& prevention},
  volume={92},
  pages={230--239},
  year={2016},
  publisher={Elsevier}
}

@article{jin2025impact,
  title={Impact of non-driving related task types, request modalities, and automation on driver takeover: A meta-analysis},
  author={Jin, Lisheng and Liu, Xingchen and Guo, Baicang and Han, Zhuotong and Wang, Yinlin and Cao, Yuan and Yang, Xiao and Shi, Jian},
  journal={Safety science},
  volume={181},
  pages={106704},
  year={2025},
  publisher={Elsevier}
}

@article{louw2017coming,
  title={Coming back into the loop: Drivers’ perceptual-motor performance in critical events after automated driving},
  author={Louw, Tyron and Markkula, Gustav and Boer, Erwin and Madigan, Ruth and Carsten, Oliver and Merat, Natasha},
  journal={Accident Analysis \& Prevention},
  volume={108},
  pages={9--18},
  year={2017},
  publisher={Elsevier}
}

@article{merat2014transition,
  title={Transition to manual: Driver behaviour when resuming control from a highly automated vehicle},
  author={Merat, Natasha and Jamson, A Hamish and Lai, Frank CH and Daly, Michael and Carsten, Oliver MJ},
  journal={Transportation research part F: traffic psychology and behaviour},
  volume={27},
  pages={274--282},
  year={2014},
  publisher={Elsevier}
}

@article{vlakveld2018situation,
  title={Situation awareness increases when drivers have more time to take over the wheel in a Level 3 automated car: A simulator study},
  author={Vlakveld, Willem and van Nes, Nicole and de Bruin, Jonathan and Vissers, Luuk and van der Kroft, Merle},
  journal={Transportation research part F: traffic psychology and behaviour},
  volume={58},
  pages={917--929},
  year={2018},
  publisher={Elsevier}
}

@article{heinrich2025ready,
  title={Ready to transition? Strategies to bring drivers back into the loop upon requests to intervene in conditionally automated driving},
  author={Heinrich, Luisa Katharina and Wu, Yanbin and Miller, Linda and Metzger, Luise and Kihara, Ken and Sato, Toshihisa and Kitazaki, Satoshi and Baumann, Martin},
  journal={Transportation Research Part F: Traffic Psychology and Behaviour},
  volume={110},
  pages={163--181},
  year={2025},
  publisher={Elsevier}
}

@inproceedings{singh2025trust,
  title={Trust and Visual Focus in Automated Vehicles: A Comparative Study of Beginner and Experienced Drivers},
  author={Singh, Richa and Ziat, Mounia and {\v{S}}pakov, Oleg and M{\"a}kel{\"a}, John and Surakka, Veikko and Raisamo, Roope},
  booktitle={Proceedings of the 2025 CHI Conference on Human Factors in Computing Systems},
  pages={1--11},
  year={2025}
}

@article{pan2023impact,
  title={The impact of non-driving related tasks on the development of driver sleepiness and takeover performances in prolonged automated driving},
  author={Pan, Hengyan and He, Haijing and Wang, Yonggang and Cheng, Yanqiu and Dai, Zhe},
  journal={Journal of safety research},
  volume={86},
  pages={148--163},
  year={2023},
  publisher={Elsevier}
}

@article{hu2024non,
  title={Non-driving-related tasks and drivers’ takeover time: A meta-analysis},
  author={Hu, Wenbo and Zhang, Tingru and Zhang, Yan and Chan, Alan Hoi Shou},
  journal={Transportation research part F: traffic psychology and behaviour},
  volume={103},
  pages={623--637},
  year={2024},
  publisher={Elsevier}
}

@article{wu2025effect,
  title={Effect of takeover request time and warning modality on trust in L3 automated driving},
  author={Wu, Yu and Yao, Xiaoyu and Deng, Fenghui and Yuan, Xiaofang},
  journal={Human Factors},
  volume={67},
  number={5},
  pages={427--444},
  year={2025},
  publisher={SAGE Publications Sage CA: Los Angeles, CA}
}

@article{chai2024effects,
  title={The effects of various auditory takeover requests: A simulated driving study considering the modality of non-driving-related tasks},
  author={Chai, Chunlei and Lei, Yu and Wei, Haoran and Wu, Changxu and Zhang, Wei and Hansen, Preben and Fan, Hao and Shi, Jinlei},
  journal={Applied Ergonomics},
  volume={118},
  pages={104252},
  year={2024},
  publisher={Elsevier}
}

@article{dong2026multimodal,
  title={Multimodal attention-guiding takeover alerts in autonomous driving: Comparing visual, auditory, and combined modalities},
  author={Dong, Zhanxun and Li, Yahong and Fang, Zhou and Xu, Heyi and Putri, Yessy Liana},
  journal={International Journal of Industrial Ergonomics},
  volume={112},
  pages={103879},
  year={2026},
  publisher={Elsevier}
}

@inproceedings{srour2025all,
  title={All-inclusive TORs: Cross-Cultural and Age-Sensitive Design for Take-Over Requests in Level 3 Cars},
  author={Srour Zreik, Rawan and Harvey, Monika and Brewster, Stephen Anthony},
  booktitle={Proceedings of the 2025 CHI Conference on Human Factors in Computing Systems},
  pages={1--17},
  year={2025}
}

@article{hong2022effect,
  title={Effect of multimodal takeover request issued through A-pillar LED light, earcon, speech message, and haptic seat in conditionally automated driving},
  author={Hong, Sara and Yang, Ji Hyun},
  journal={Transportation research part F: traffic psychology and behaviour},
  volume={89},
  pages={488--500},
  year={2022},
  publisher={Elsevier}
}

@article{rodak2025hmi,
  title={HMI efficiency, usability and workload during take-over in AVs},
  author={Rodak, Aleksandra and Kruszewski, Miko{\l}aj and Niedzicka, Anna},
  journal={Scientific Reports},
  volume={15},
  number={1},
  pages={15176},
  year={2025},
  publisher={Nature Publishing Group UK London}
}

@article{srour2024s,
  title={Where's my TOR?: Evaluating the Effect of Take-Over Request Source on Older Drivers' Control Transition in Level 3 Cars},
  author={Srour Zreik, Rawan and Goodge, Thomas and Harvey, Monika and Brewster, Stephen},
  journal={Proceedings of the ACM on Human-Computer Interaction},
  volume={8},
  number={MHCI},
  pages={1--25},
  year={2024},
  publisher={ACM New York, NY, USA}
}

@article{yu2025impact,
  title={Impact of information-assisted modalities and levels on driver takeover performance in conditional automated driving},
  author={Yu, Zhenhua and Zhai, Linjing and Jiang, Kang and Yu, Shuangyu and Zhang, Bingzhan and Deng, Qingqing and Huang, Zhipeng},
  journal={Accident Analysis \& Prevention},
  volume={220},
  pages={108158},
  year={2025},
  publisher={Elsevier}
}

@article{zhang2025effects,
  title={Effects of time interval and request modality on driver takeover responses: Identifying the optimal time interval for two-stage warning system},
  author={Zhang, Jie and Zhang, Zhi and Zhang, Tingru and Zhang, Yijing and Chen, Shanguang},
  journal={Accident Analysis \& Prevention},
  volume={215},
  pages={108008},
  year={2025},
  publisher={Elsevier}
}

@inproceedings{sadeghian2018feel,
  title={Feel the movement: Real motion influences responses to take-over requests in highly automated vehicles},
  author={Sadeghian Borojeni, Shadan and Boll, Susanne CJ and Heuten, Wilko and B{\"u}lthoff, Heinrich H and Chuang, Lewis},
  booktitle={Proceedings of the 2018 CHI Conference on Human Factors in Computing Systems},
  pages={1--13},
  year={2018}
}

@article{wang2024human,
  title={Human-Machine interfaces and vehicle Automation: A review of the literature and recommendations for system Design, Feedback, and alerts},
  author={Wang, Meng and Mehrotra, Shashank and Wong, Nicholas and Parker, Jah’inaya and Roberts, Shannon C and Kim, Woon and Romo, Alicia and Horrey, William J},
  journal={Transportation Research Part F: Traffic Psychology and Behaviour},
  volume={107},
  pages={549--561},
  year={2024},
  publisher={Elsevier}
}

@article{shi2026striving,
  title={Striving for a balance between safety performance and driver acceptance in auditory takeover requests},
  author={Shi, Jinlei and Lei, Yu and Zhang, Wei and Yin, Min and Fan, Hao and Chai, Chunlei},
  journal={Applied Ergonomics},
  volume={131},
  pages={104668},
  year={2026},
  publisher={Elsevier}
}

@article{huang2022takeover,
  title={Takeover requests for automated driving: The effects of signal direction, lead time, and modality on takeover performance},
  author={Huang, Gaojian and Pitts, Brandon J},
  journal={Accident Analysis \& Prevention},
  volume={165},
  pages={106534},
  year={2022},
  publisher={Elsevier}
}

@article{bazilinskyy2018take,
  title={Take-over requests in highly automated driving: A crowdsourcing survey on auditory, vibrotactile, and visual displays},
  author={Bazilinskyy, Pavlo and Petermeijer, Sebastiaan M and Petrovych, Veronika and Dodou, Dimitra and de Winter, Joost CF},
  journal={Transportation research part F: traffic psychology and behaviour},
  volume={56},
  pages={82--98},
  year={2018},
  publisher={Elsevier}
}

@article{pan2024age,
  title={Age-related differences in takeover performance: A comparative analysis of older and younger drivers in prolonged partially automated driving},
  author={Pan, Hengyan and Payre, William and Xu, Jinhua and Koppel, Sjaan},
  journal={Traffic injury prevention},
  volume={25},
  number={7},
  pages={968--975},
  year={2024},
  publisher={Taylor \& Francis}
}

@article{yin2024monitoring,
  title={The monitoring requests on young driver’s fatigue and take-over performance in prolonged conditional automated driving},
  author={Yin, Juan and Shao, Haipeng and Zhang, Xinjie},
  journal={Journal of safety research},
  volume={88},
  pages={285--292},
  year={2024},
  publisher={Elsevier}
}

@article{coyne2026effect,
  title={The effect of prolonged conditionally automated driving on fatigue, physiological activity, and takeover performance: a driving simulator experiment},
  author={Coyne, Rory and Khan, Bilal and Shiel, Matthew and Rahman, Shams Ur and Vlad, Victor and Lillis, Conor and Usman, Muzna and Kielty, Paul and Parsi, Ashkan and Lemley, Joseph and others},
  journal={Physiological Measurement},
  volume={47},
  number={2},
  pages={025013},
  year={2026},
  publisher={IOP Publishing}
}

@article{loew2025impact,
  title={Impact of free-choice non-driving related activities on driver fatigue and performance in the context of conditional automated driving},
  author={Loew, Alexandra and Kurpiers, Christina and G{\"o}tze, Martin and Gold, Christian and Bengler, Klaus},
  journal={Transportation Research Part F: Traffic Psychology and Behaviour},
  volume={114},
  pages={216--229},
  year={2025},
  publisher={Elsevier}
}

@article{ma2026video,
  title={Video entertainment to mitigate safety risks during prolonged automated driving: Impacts on cognitive load, fatigue, and takeover performance},
  author={Ma, Jun and Zang, Lizhuo and Li, Jiateng and Gong, Zaiyan and Fei, Yue and Lu, Jin},
  journal={Applied Ergonomics},
  volume={130},
  pages={104644},
  year={2026},
  publisher={Elsevier}
}

@article{liu2024safety,
  title={Safety evaluation and prediction of takeover performance in automated driving considering drivers’ cognitive load: A driving simulator study},
  author={Liu, Yongjie and Wu, Chaozhong and Zhang, Hui and Ding, Naikan and Xiao, Yiying and Zhang, Qi and Tian, Kai},
  journal={Transportation research part F: traffic psychology and behaviour},
  volume={103},
  pages={35--52},
  year={2024},
  publisher={Elsevier}
}

@inproceedings{sun2026adaptive,
  title={Adaptive Bounded-Rationality Modeling of Early-Stage Takeover in Shared-Control Driving},
  author={Sun, Jian and Jiang, Xiyan and Zhao, Xiaocong and Wang, Jie and Hang, Peng and Li, Zirui},
  booktitle={Proceedings of the 2026 CHI Conference on Human Factors in Computing Systems},
  pages={1--23},
  year={2026}
}

@inproceedings{sheem2026must,
  title={MUST: Smartwatch-based Multimodal Framework for Predicting Driver State and Takeover Performance},
  author={Sheem, Seokyong and Cho, Yujin and Lee, In Kyung and Cho, Hanjun and Kim, Taegeun and Kim, Byung Hyung and Kim, Daekyum},
  booktitle={Proceedings of the 2026 CHI Conference on Human Factors in Computing Systems},
  pages={1--17},
  year={2026}
}

@article{zhang2025objective,
  title={Objective evaluation of 35 heart rate variability parameters for predicting takeover performance in conditionally automated driving},
  author={Zhang, N and Tohmuang, S and Fard, M and Davy, JL and Robinson, SR},
  journal={International Journal of Industrial Ergonomics},
  volume={106},
  pages={103699},
  year={2025},
  publisher={Elsevier}
}

@inproceedings{kaufman2025did,
  title={What did my car say? impact of autonomous vehicle explanation errors and driving context on comfort, reliance, satisfaction, and driving confidence},
  author={Kaufman, Robert A and Broukhim, Aaron and Kirsh, David and Weibel, Nadir},
  booktitle={Proceedings of the 2025 CHI Conference on Human Factors in Computing Systems},
  pages={1--17},
  year={2025}
}

@article{tan2025driver,
  title={Driver situation awareness for regaining control from conditionally automated vehicles: A systematic review of empirical studies},
  author={Tan, Xiaomei and Zhang, Yiqi},
  journal={Human factors},
  volume={67},
  number={4},
  pages={367--403},
  year={2025},
  publisher={SAGE Publications Sage CA: Los Angeles, CA}
}

@inproceedings{zhang2025actual,
  title={Actual Achieved Gain and Optimal Perceived Gain: Modeling Human Take-over Decisions Towards Automated Vehicles' Suggestions},
  author={Zhang, Shuning and Yi, Xin and Li, Shixuan and Hong, Chuye and Chen, Gujun and Liu, Jiarui and Wang, Xueyang and Hu, Yongquan and Wang, Yuntao and Li, Hewu},
  booktitle={Proceedings of the 2025 CHI Conference on Human Factors in Computing Systems},
  pages={1--18},
  year={2025}
}

@article{van2016supporting,
  title={Supporting the changing driver’s task: Exploration of interface designs for supervision and intervention in automated driving},
  author={Van den Beukel, Arie P and van der Voort, Mascha C and Eger, Arthur O},
  journal={Transportation research part F: traffic psychology and behaviour},
  volume={43},
  pages={279--301},
  year={2016},
  publisher={Elsevier}
}

@incollection{walch2017car,
  title={From car-driver-handovers to cooperative interfaces: Visions for driver--vehicle interaction in automated driving},
  author={Walch, Marcel and M{\"u}hl, Kristin and Kraus, Johannes and Stoll, Tanja and Baumann, Martin and Weber, Michael},
  booktitle={Automotive user interfaces: Creating interactive experiences in the car},
  pages={273--294},
  year={2017},
  publisher={Springer}
}

@article{zhang2026novel,
  title={A novel driver monitoring and feedback system improves takeover performance in conditional automated driving},
  author={Zhang, N and Fard, M and Tohmuang, S and Xu, J and Davy, JL and Robinson, SR},
  journal={International Journal of Industrial Ergonomics},
  volume={111},
  pages={103862},
  year={2026},
  publisher={Elsevier}
}

@article{wang2025exploring,
  title={Exploring how physio-psychological states affect drivers’ takeover performance in conditional automated vehicles},
  author={Wang, Ange and Wang, Jiyao and Huang, Chunxi and He, Dengbo and Yang, Hai},
  journal={Accident Analysis \& Prevention},
  volume={216},
  pages={108022},
  year={2025},
  publisher={Elsevier}
}

@inproceedings{wintersberger2018let,
  title={Let me finish before I take over: Towards attention aware device integration in highly automated vehicles},
  author={Wintersberger, Philipp and Riener, Andreas and Schartm{\"u}ller, Clemens and Frison, Anna-Katharina and Weigl, Klemens},
  booktitle={Proceedings of the 10th international conference on automotive user interfaces and interactive vehicular applications},
  pages={53--65},
  year={2018}
}

@article{weaver2022systematic,
  title={A systematic review and meta-analysis of takeover performance during conditionally automated driving},
  author={Weaver, Bradley W and DeLucia, Patricia R},
  journal={Human factors},
  volume={64},
  number={7},
  pages={1227--1260},
  year={2022},
  publisher={Sage Publications Sage CA: Los Angeles, CA}
}

@article{deng2024design,
  title={How to design driver takeover request in real-world scenarios: A systematic review},
  author={Deng, Hanwen and Xiang, Guoliang and Pan, Jiandong and Wu, Xianhui and Fan, Chaojie and Wang, Kui and Peng, Yong},
  journal={Transportation Research Part F: Traffic Psychology and Behaviour},
  volume={104},
  pages={411--432},
  year={2024},
  publisher={Elsevier}
}

@article{li2025obstacle,
  title={How do obstacle characteristics and driver alertness affect the takeover process in conditionally automated driving?},
  author={Li, Penghui and Zhu, Shufen and Zhang, Ni and Gong, Jianguo and Li, Xiaomeng and Dong, Chunjiao and Yan, Xuedong},
  journal={Transportation Research Part F: Traffic Psychology and Behaviour},
  volume={114},
  pages={1253--1267},
  year={2025},
  publisher={Elsevier}
}

@inproceedings{morgan2017manual,
  title={Manual takeover and handover of a simulated fully autonomous vehicle within urban and extra-urban settings},
  author={Morgan, Phillip L and Alford, Chris and Williams, Craig and Parkhurst, Graham and Pipe, Tony},
  booktitle={International Conference on Applied Human Factors and Ergonomics},
  pages={760--771},
  year={2017},
  organization={Springer}
}

@article{guo2024analysis,
  title={Analysis of influencing factors of level 3 automated vehicle takeover: a literature review},
  author={Guo, Hanying and Qiu, Haoyu and Zhou, Yongjiang and Deng, Yuxin},
  journal={Sustainability},
  volume={16},
  number={19},
  pages={8345},
  year={2024},
  publisher={MDPI}
}

@article{may2009driver,
  title={Driver fatigue: The importance of identifying causal factors of fatigue when considering detection and countermeasure technologies},
  author={May, Jennifer F and Baldwin, Carryl L},
  journal={Transportation research part F: traffic psychology and behaviour},
  volume={12},
  number={3},
  pages={218--224},
  year={2009},
  publisher={Elsevier}
}

@inproceedings{saxby2007development,
  title={Development of active and passive fatigue manipulations using a driving simulator},
  author={Saxby, Dyani J and Matthews, Gerald and Hitchcock, Edward M and Warm, Joel S},
  booktitle={Proceedings of the Human Factors and Ergonomics Society Annual Meeting},
  volume={51},
  number={18},
  pages={1237--1241},
  year={2007},
  organization={SAGE Publications Sage CA: Los Angeles, CA}
}

@article{matthews2002task,
  title={Task-induced fatigue states and simulated driving performance},
  author={Matthews, Gerald and Desmond, Paula A},
  journal={The Quarterly Journal of Experimental Psychology: Section A},
  volume={55},
  number={2},
  pages={659--686},
  year={2002},
  publisher={Taylor \& Francis}
}

@article{thiffault2003monotony,
  title={Monotony of road environment and driver fatigue: a simulator study},
  author={Thiffault, Pierre and Bergeron, Jacques},
  journal={Accident Analysis \& Prevention},
  volume={35},
  number={3},
  pages={381--391},
  year={2003},
  publisher={Elsevier}
}

@article{aakerstedt1990subjective,
  title={Subjective and objective sleepiness in the active individual},
  author={{\AA}kerstedt, Torbj{\"o}rn and Gillberg, Mats},
  journal={International journal of neuroscience},
  volume={52},
  number={1-2},
  pages={29--37},
  year={1990},
  publisher={Taylor \& Francis}
}

@article{kaida2006validation,
  title={Validation of the Karolinska sleepiness scale against performance and EEG variables},
  author={Kaida, Kosuke and Takahashi, Masaya and {\AA}kerstedt, Torbj{\"o}rn and Nakata, Akinori and Otsuka, Yasumasa and Haratani, Takashi and Fukasawa, Kenji},
  journal={Clinical neurophysiology},
  volume={117},
  number={7},
  pages={1574--1581},
  year={2006},
  publisher={Elsevier}
}

@article{kennedy1993simulator,
  title={Simulator sickness questionnaire: An enhanced method for quantifying simulator sickness},
  author={Kennedy, Robert S and Lane, Norman E and Berbaum, Kevin S and Lilienthal, Michael G},
  journal={The international journal of aviation psychology},
  volume={3},
  number={3},
  pages={203--220},
  year={1993},
  publisher={Taylor \& Francis}
}

@article{kaptein1996driving,
  title={Driving simulator validity: Some considerations},
  author={Kaptein, Nico A and Theeuwes, Jan and Van Der Horst, Richard},
  journal={Transportation research record},
  volume={1550},
  number={1},
  pages={30--36},
  year={1996},
  publisher={SAGE Publications Sage CA: Los Angeles, CA}
}

@inproceedings{li2025researchers,
  title={What Researchers Need from Driving Simulator Systems: A Thematic Analysis of Expert Interviews},
  author={Li, Stacey and Dey, Debargha and Santacruz, Claudia and Ju, Wendy},
  booktitle={Proceedings of the 17th International Conference on Automotive User Interfaces and Interactive Vehicular Applications},
  pages={55--68},
  year={2025}
}

@inproceedings{dosovitskiy2017carla,
  title={CARLA: An open urban driving simulator},
  author={Dosovitskiy, Alexey and Ros, German and Codevilla, Felipe and Lopez, Antonio and Koltun, Vladlen},
  booktitle={Conference on robot learning},
  pages={1--16},
  year={2017},
  organization={PMLR}
}

@article{hayward1972near,
  title={Near miss determination through use of a scale of danger},
  author={Hayward, John C},
  year={1972},
  publisher={Pennsylvania State University University Park, PA, USA}
}

@inproceedings{obrist2014opportunities,
  title={Opportunities for odor: experiences with smell and implications for technology},
  author={Obrist, Marianna and Tuch, Alexandre N and Hornbaek, Kasper},
  booktitle={Proceedings of the SIGCHI Conference on Human Factors in Computing Systems},
  pages={2843--2852},
  year={2014}
}

@inproceedings{brewster2006olfoto,
  title={Olfoto: designing a smell-based interaction},
  author={Brewster, Stephen and McGookin, David and Miller, Christopher},
  booktitle={Proceedings of the SIGCHI conference on Human Factors in computing systems},
  pages={653--662},
  year={2006}
}

@inproceedings{fu2024aromablendz,
  title={Aromablendz: an olfactory system for crafting personalized scents},
  author={Fu, Shihan and Chen, Jianhao and Cai, Yi and Fan, Mingming},
  booktitle={Extended Abstracts of the CHI Conference on Human Factors in Computing Systems},
  pages={1--5},
  year={2024}
}

@article{liu2024intelligent,
  title={Intelligent wearable olfactory interface for latency-free mixed reality and fast olfactory enhancement},
  author={Liu, Yiming and Jia, Shengxin and Yiu, Chun Ki and Park, Wooyoung and Chen, Zhenlin and Nan, Jin and Huang, Xingcan and Chen, Hongting and Li, Wenyang and Gao, Yuyu and others},
  journal={Nature Communications},
  volume={15},
  number={1},
  pages={4474},
  year={2024},
  publisher={Nature Publishing Group UK London}
}

@article{spence2017digitizing,
  title={Digitizing the chemical senses: Possibilities \& pitfalls},
  author={Spence, Charles and Obrist, Marianna and Velasco, Carlos and Ranasinghe, Nimesha},
  journal={International Journal of Human-Computer Studies},
  volume={107},
  pages={62--74},
  year={2017},
  publisher={Elsevier}
}

@article{maggioni2020smell,
  title={SMELL SPACE: mapping out the olfactory design space for novel interactions},
  author={Maggioni, Emanuela and Cobden, Robert and Dmitrenko, Dmitrijs and Hornb{\ae}k, Kasper and Obrist, Marianna},
  journal={ACM Transactions on Computer-Human Interaction (TOCHI)},
  volume={27},
  number={5},
  pages={1--26},
  year={2020},
  publisher={ACM New York, NY, USA}
}

@inproceedings{chang2025crossmodal,
  title={Crossmodal interactions in human-robot communication: Exploring the influences of scent and voice congruence on user perceptions of social robots},
  author={Chang, Fangyuan and Chen, Bingliang and Zhang, Xingguo and Sheng, Lin and Zhu, Dian and Zhao, Jianan and Gu, Zhenyu},
  booktitle={Proceedings of the 2025 CHI Conference on Human Factors in Computing Systems},
  pages={1--15},
  year={2025}
}

@inproceedings{dmitrenko2016comparison,
  title={A comparison of scent-delivery devices and their meaningful use for in-car olfactory interaction},
  author={Dmitrenko, Dmitrijs and Vi, Chi Thanh and Obrist, Marianna},
  booktitle={Proceedings of the 8th International Conference on Automotive User Interfaces and Interactive Vehicular Applications},
  pages={23--26},
  year={2016}
}

@article{wang2025multi,
  title={Multi-scent olfactory display with visual and sound interaction},
  author={Wang, Chuhong and Daulton, Emma and Covington, James A},
  journal={International Journal of Human-Computer Studies},
  volume={197},
  pages={103421},
  year={2025},
  publisher={Elsevier}
}

@inproceedings{dmitrenko2017did,
  title={What did I sniff? Mapping scents onto driving-related messages},
  author={Dmitrenko, Dmitrijs and Maggioni, Emanuela and Vi, Chi Thanh and Obrist, Marianna},
  booktitle={Proceedings of the 9th international conference on automotive user interfaces and interactive vehicular applications},
  pages={154--163},
  year={2017}
}

@inproceedings{dmitrenko2018smell,
  title={I smell trouble: using multiple scents to convey driving-relevant information},
  author={Dmitrenko, Dmitrijs and Maggioni, Emanuela and Obrist, Marianna},
  booktitle={Proceedings of the 20th ACM international conference on multimodal interaction},
  pages={234--238},
  year={2018}
}

@inproceedings{brooks2023third,
  title={Third wave or winter? The past and future of smell in HCI},
  author={Brooks, Jas and Lopes, Pedro and Obrist, Marianna and Amores Fernandez, Judith and Kaye, Jofish},
  booktitle={Extended Abstracts of the 2023 CHI Conference on Human Factors in Computing Systems},
  pages={1--4},
  year={2023}
}

@inproceedings{dmitrenko2020caroma,
  title={Caroma therapy: pleasant scents promote safer driving, better mood, and improved well-being in angry drivers},
  author={Dmitrenko, Dmitrijs and Maggioni, Emanuela and Brianza, Giada and Holthausen, Brittany E and Walker, Bruce N and Obrist, Marianna},
  booktitle={Proceedings of the 2020 chi conference on human factors in computing systems},
  pages={1--13},
  year={2020}
}

@inproceedings{schartmuller2020sick,
  title={Sick of scents: Investigating non-invasive olfactory motion sickness mitigation in automated driving},
  author={Schartm{\"u}ller, Clemens and Riener, Andreas},
  booktitle={12th International conference on automotive user interfaces and interactive vehicular applications},
  pages={30--39},
  year={2020}
}

@inproceedings{obrist2026follow,
  title={Follow Your Nose: Experiencing Smell as a Design Material},
  author={Obrist, Marianna and Be{\c{s}}evli, Ceylan and Maggioni, Emanuela and Lai, Mei-Kei and Velasco, Carlos},
  booktitle={Proceedings of the Extended Abstracts of the 2026 CHI Conference on Human Factors in Computing Systems},
  pages={1--3},
  year={2026}
}

@inproceedings{alkurdi2025trust,
  title={Trust in Level 3 automated vehicles: A systematic literature review},
  author={Alkurdi, Duha and Alsaid, Areen},
  booktitle={Proceedings of the Human Factors and Ergonomics Society Annual Meeting},
  volume={69},
  number={1},
  pages={1281--1283},
  year={2025},
  organization={SAGE Publications Sage CA: Los Angeles, CA}
}

@article{tang2021olfactory,
  title={Olfactory facilitation of takeover performance in highly automated driving},
  author={Tang, Qiuyang and Guo, Gang and Zhang, Zijian and Zhang, Bingbing and Wu, Yingzhang},
  journal={Human factors},
  volume={63},
  number={4},
  pages={553--564},
  year={2021},
  publisher={Sage Publications Sage CA: Los Angeles, CA}
}

\appendix
\newpage

\section{Pilot Peppermint Takeover-Cue Questionnaire}
\label{app:pilot-questionnaire}

The pilot questionnaire was administered to evaluate the detectability, perceived intensity, comfort, alerting experience, suitability, familiarity, and acceptance of candidate peppermint cues. All rating items used 1--7 scales.

The complete English questionnaire items and scale anchors are reproduced in Table~\ref{tab:pilot-questionnaire}.

\begin{table*}[!htbp]
\centering
\caption{The pilot peppermint questionnaire.}
\label{tab:pilot-questionnaire}
\small
\renewcommand{\arraystretch}{1.25}
\setlength{\tabcolsep}{4pt}

\begin{tabular}{
c
p{0.57\textwidth}
p{0.32\textwidth}
}
\hline
\textbf{Item} &
\textbf{Question} &
\textbf{Scale anchors} \\
\hline

1 &
I could clearly detect the peppermint odor. &
1 = Not detectable at all; 7 = Very easy to detect. \\

2 &
The overall intensity of the peppermint odor was appropriate. &
1 = Very weak; 7 = Very strong. \\

3 &
I found the peppermint odor comfortable overall. &
1 = Very uncomfortable; 7 = Very comfortable. \\

4 &
The peppermint odor produced a noticeable stimulating sensation. &
1 = Not noticeable at all; 7 = Very noticeable. \\

5 &
The peppermint odor made me feel more alert. &
1 = Not at all; 7 = Very noticeably. \\

6 &
I think the peppermint odor is suitable as an automated-driving takeover reminder. &
1 = Completely unsuitable; 7 = Very suitable. \\

7 &
I am familiar with the peppermint odor. &
1 = Strongly disagree; 7 = Strongly agree. \\

8 &
If a future vehicle used this type of reminder, I would be willing to accept it. &
1 = Completely unwilling; 7 = Very willing. \\

\hline
\end{tabular}
\end{table*}

Participants also answered the following post-comparison questions:

\begin{enumerate}
    \item Which odor condition do you think is most suitable as an automated-driving takeover reminder? (A / B / C / D)
    \item Why did you select this condition?
    \item Do you have any additional suggestions for using peppermint as a takeover reminder?
\end{enumerate}

The experimenter additionally recorded the actual concentration, spray duration, randomized presentation order, and relevant notes for each candidate condition.


\section{Post-Study Semi-Structured Interview Guide}
\label{app:interview-guide}

The interview began with the following neutral instruction:

\begin{quote}
We would now like to briefly understand your actual experience of the three takeover-reminder approaches. There are no correct or incorrect answers. Please answer according to what you genuinely experienced. We are interested in why you felt this way, rather than evaluating your driving performance.
\end{quote}

The interview used the following core questions and optional probes:

\begin{enumerate}
    \item \textbf{Overall comparison.} Thinking back across the three experimental blocks, how did the three takeover reminders feel different to you?
    \begin{itemize}
        \item Which one was most memorable, and why?
        \item What specifically did you feel?
    \end{itemize}

    \item \textbf{Audio-only (S1).} In the condition with only the auditory reminder, how did you feel after hearing the sound?
    \begin{itemize}
        \item Was the reminder clear to you?
        \item Did you immediately understand that you needed to take over the vehicle?
        \item Did it feel comfortable or sudden? Why?
        \item Would you accept this type of reminder in future use?
    \end{itemize}

    \item \textbf{Aligned (S2) olfactory--auditory reminder.} When the sound and odor were initiated together, did you notice the sound, the odor, or both?
    \begin{itemize}
        \item Did adding the odor change how the takeover reminder felt?
        \item Did it make the reminder clearer, or did it feel redundant or distracting?
    \end{itemize}

    \item \textbf{Cue-Lead (S3) olfactory-to-auditory reminder.} When the odor appeared before the sound, what did you feel when you first noticed the odor? What did you think the odor meant at that moment?
    \begin{itemize}
        \item Did you expect that something might happen next?
        \item After noticing the odor, did you pay more attention to the road, or did you remain in the same state?
        \item When the sound appeared later, how did that experience differ from the other two reminders?
    \end{itemize}

    \item \textbf{Comfort and annoyance.} Was any reminder particularly comfortable, uncomfortable, sudden, or annoying?

    \item \textbf{Perceived effectiveness.} From your own perspective, which reminder helped you most with the takeover?
    \begin{itemize}
        \item How did it help?
        \item Did it make the reminder easier to notice, easier to understand, or make you feel more prepared?
        \item Was any reminder very noticeable but not necessarily preferable?
    \end{itemize}

    \item \textbf{Long-term acceptance.} If these reminders were used in a vehicle you drove regularly, which one would you most prefer?
    \begin{itemize}
        \item Would you have any concerns about frequent use of odor reminders?
        \item If you encountered such reminders many times per day, do you think your experience would change?
    \end{itemize}

    \item \textbf{Peppermint experience.} Overall, what did you think of the peppermint odor?
    \begin{itemize}
        \item Was it easy to detect?
        \item How was the intensity?
        \item Was it comfortable or uncomfortable?
        \item Did it feel too long, too short, or remain after the experimental event?
    \end{itemize}

    \item \textbf{Participant-designed reminder.} If you could design an automated-driving takeover reminder yourself, what approach would you choose? What would you improve?

    \item \textbf{Open comments.} Is there anything else you would like to add or suggest?
\end{enumerate}


\section{Dataset Availability and Quality Audit}
\label{app:data-audit}

The intended protocol contained three condition slots per participant. Across 18 participants this corresponds to 54 planned slots; 33 trial rows are present and 21 planned slots did not yield analyzable trial records because of equipment-related technical interruptions during data collection. For anonymous reporting, the 18 analyzed participants are relabeled consecutively P01--P18 in Table~\ref{tab:condition-slot-flow}; this presentation-only relabeling preserves the one-to-one participant/trial mapping and does not alter the underlying data. Detailed per-block fault categories were not retained, so the table records availability without attributing each interruption to a specific subsystem.

\begin{table}[!htbp]
\centering
\caption{Condition-slot provenance by consecutively relabeled participants P01--P18. Only P01 and P02 contain all three planned conditions. Unavailable slots resulted from equipment-related technical interruptions; detailed per-block fault categories were not retained.}
\label{tab:condition-slot-flow}
\small
\renewcommand{\arraystretch}{1.18}
\setlength{\tabcolsep}{3pt}

\begin{tabular}{
c
p{0.25\columnwidth}
p{0.31\columnwidth}
c
}
\hline
\textbf{Session} &
\textbf{Available conditions} &
\textbf{Unavailable conditions} &
\textbf{Trial rows} \\
\hline

P01 & S1, S2, S3 & --- & 3 \\
P02 & S1, S2, S3 & --- & 3 \\
P03 & S1 & S2, S3 & 1 \\
P04 & S2, S3 & S1 & 2 \\
P05 & S3 & S1, S2 & 1 \\
P06 & S1, S2 & S3 & 2 \\
P07 & S1, S2 & S3 & 2 \\
P08 & S1, S2 & S3 & 2 \\
P09 & S1, S2 & S3 & 2 \\
P10 & S2, S3 & S1 & 2 \\
P11 & S1, S2 & S3 & 2 \\
P12 & S1, S2 & S3 & 2 \\
P13 & S1, S2 & S3 & 2 \\
P14 & S3 & S1, S2 & 1 \\
P15 & S1, S2 & S3 & 2 \\
P16 & S1, S2 & S3 & 2 \\
P17 & S3 & S1, S2 & 1 \\
P18 & S3 & S1, S2 & 1 \\

\hline
\end{tabular}
\end{table}

Condition-level coverage is S1: 12/18 participants, S2: 13/18, and S3: 8/18. Although the unavailable slots are attributed to technical interruptions rather than outcome-based exclusion, the unequal pattern does not establish that interruption probability was independent of condition. The main text therefore retains a conservative interpretation of the observed condition means.

Metric-level missingness among the 33 available trial rows is summarized in Table~\ref{tab:data-availability}. No available trial row was removed because of an outlier flag or historical pilot/FAST metadata during the final audit; missing outcome fields remained missing rather than being imputed.

\begin{table*}[t]
\centering
\caption{Availability of major variables among the 33 available trial rows.}
\label{tab:data-availability}

\small
\renewcommand{\arraystretch}{1.22}
\setlength{\tabcolsep}{4pt}

\begin{tabular}{
p{0.27\textwidth}
c
c
p{0.45\textwidth}
}
\hline
\textbf{Variable group} &
\textbf{Available} &
\textbf{Missing} &
\textbf{Note} \\
\hline

Planned condition slots &
33/54 &
21/54 &
Unavailable because of equipment-related technical interruptions; detailed fault categories not retained. \\

Takeover Time &
26/33 &
7 &
Metric-specific missingness retained. \\

Collision &
33/33 &
0 &
Complete for available trials. \\

Minimum TTC &
33/33 &
0 &
Complete for available trials. \\

Maximum deceleration &
33/33 &
0 &
Complete for available trials. \\

Lane-center offset &
33/33 &
0 &
Complete for available trials. \\

KSS-BlockStart / KSS-PreTOR &
33/33 &
0 &
Complete at block level. \\

Comfort / Annoyance / Acceptance / Perceived Effectiveness &
30/33 &
3 &
Primary formal subjective outcomes. \\

Odor Detection / Intensity &
30/33 &
3 &
Descriptive manipulation-check information. \\

Odor Pleasantness / Arousal &
18/33 &
15 &
Available only for a subset of odor-assisted records. \\

Residual &
0/33 &
33 &
No analyzable numerical Residual field in the frozen export. \\

\hline
\end{tabular}
\end{table*}

\section{Supplementary Descriptive Statistics}
\label{app:descriptive-statistics}

Tables in this appendix provide the exact values underlying the main-text result figures. Percentile bootstrap 95\% confidence intervals were computed from 10,000 within-condition resamples of the available session-level observations (fixed random seed 2027). They are descriptive uncertainty summaries conditional on the observed data; they do not correct technical-interruption-related condition imbalance or the condition--scenario confound.

\begin{table*}[t]
\centering
\caption{Behavioral and safety outcomes by reminder condition. Brackets show descriptive bootstrap 95\% confidence intervals for the mean/proportion.}
\label{tab:supp-behavior}

\scriptsize
\renewcommand{\arraystretch}{1.30}
\setlength{\tabcolsep}{3.5pt}

\begin{tabular}{
p{0.21\textwidth}
p{0.245\textwidth}
p{0.245\textwidth}
p{0.245\textwidth}
}
\hline
\textbf{Measure} &
\textbf{Audio-only (S1)} &
\textbf{Aligned (S2)} &
\textbf{Cue-Lead (S3)} \\
\hline

Takeover Time (s) &
$3.78\pm2.68$, $n=9$ [2.19, 5.48] &
$3.18\pm1.92$, $n=11$ [2.11, 4.25] &
$2.12\pm0.96$, $n=6$ [1.42, 2.80] \\

Minimum TTC (s) &
$2.36\pm2.47$, $n=12$ [1.10, 3.69] &
$1.88\pm1.99$, $n=13$ [0.89, 2.99] &
$2.98\pm2.11$, $n=8$ [1.58, 4.25] \\

Collision, $n/N$ (\%) &
6/12 (50.0\%) [25.0, 75.0] &
7/13 (53.8\%) [30.8, 76.9] &
3/8 (37.5\%) [12.5, 75.0] \\

Maximum deceleration (m/s$^2$) &
$-5.53\pm3.91$ [$-7.60$, $-3.41$] &
$-6.48\pm4.09$ [$-8.57$, $-4.25$] &
$-4.34\pm3.87$ [$-6.82$, $-1.86$] \\

Maximum lane-center offset (m) &
$0.57\pm0.82$ [0.16, 1.04] &
$1.58\pm1.33$ [0.90, 2.27] &
$1.73\pm1.20$ [0.94, 2.52] \\

\hline
\end{tabular}
\end{table*}

\begin{table*}[t]
\centering
\caption{Post-block subjective ratings (1--7). Brackets show descriptive bootstrap 95\% confidence intervals for the mean.}
\label{tab:supp-subjective}

\scriptsize
\renewcommand{\arraystretch}{1.30}
\setlength{\tabcolsep}{3.5pt}

\begin{tabular}{
p{0.21\textwidth}
p{0.245\textwidth}
p{0.245\textwidth}
p{0.245\textwidth}
}
\hline
\textbf{Measure} &
\textbf{Audio-only (S1)} &
\textbf{Aligned (S2)} &
\textbf{Cue-Lead (S3)} \\
\hline

Comfort &
$3.33\pm0.98$, $n=12$ [2.83, 3.83] &
$4.27\pm1.42$, $n=11$ [3.36, 5.00] &
$4.43\pm0.79$, $n=7$ [4.00, 5.00] \\

Annoyance &
$4.08\pm1.08$, $n=12$ [3.50, 4.67] &
$3.09\pm1.14$, $n=11$ [2.45, 3.73] &
$3.86\pm0.90$, $n=7$ [3.14, 4.43] \\

Acceptance &
$4.67\pm1.23$, $n=12$ [4.00, 5.33] &
$4.64\pm1.50$, $n=11$ [3.73, 5.36] &
$4.57\pm0.98$, $n=7$ [4.00, 5.14] \\

Perceived Effectiveness &
$4.08\pm1.24$, $n=12$ [3.42, 4.75] &
$5.09\pm1.45$, $n=11$ [4.27, 5.82] &
$4.57\pm1.13$, $n=7$ [4.00, 5.43] \\

\hline
\end{tabular}
\end{table*}

Because categorical detection responses and numerical odor ratings were not fully aligned across retained records, olfactory questionnaire values are reported only as a descriptive audit in Table~\ref{tab:odor-audit}.

\begin{table*}[t]
\centering
\caption{Descriptive audit of olfactory questionnaire responses. These values are treated as manipulation-check information rather than primary outcomes.}
\label{tab:odor-audit}

\small
\renewcommand{\arraystretch}{1.30}
\setlength{\tabcolsep}{4pt}

\begin{tabular}{
p{0.31\textwidth}
p{0.205\textwidth}
p{0.205\textwidth}
p{0.205\textwidth}
}
\hline
\textbf{Measure} &
\textbf{Audio-only (S1)} &
\textbf{Aligned (S2)} &
\textbf{Cue-Lead (S3)} \\
\hline

Detection: Yes / No / Uncertain / Missing &
0 / 11 / 1 / 0 &
8 / 2 / 1 / 2 &
2 / 5 / 0 / 1 \\

Odor intensity, mean $\pm$ SD &
$2.17\pm1.80$ ($n=12$) &
$4.64\pm1.21$ ($n=11$) &
$4.57\pm1.13$ ($n=7$) \\

Odor pleasantness, mean $\pm$ SD &
--- &
$4.45\pm1.04$ ($n=11$) &
$4.14\pm0.90$ ($n=7$) \\

Odor arousal, mean $\pm$ SD &
--- &
$4.91\pm1.04$ ($n=11$) &
$4.43\pm0.79$ ($n=7$) \\

\hline
\end{tabular}
\end{table*}

Finally, KSS state monitoring is summarized in Table~\ref{tab:kss-audit}. These values document change from block start to pre-TOR within the available condition subsets and are not used for between-condition fatigue inference.

\begin{table*}[t]
\centering
\caption{KSS state-monitoring summary by condition.}
\label{tab:kss-audit}

\small
\renewcommand{\arraystretch}{1.30}
\setlength{\tabcolsep}{4pt}

\begin{tabular}{
p{0.29\textwidth}
p{0.212\textwidth}
p{0.212\textwidth}
p{0.212\textwidth}
}
\hline
\textbf{KSS measure} &
\textbf{Audio-only (S1)} &
\textbf{Aligned (S2)} &
\textbf{Cue-Lead (S3)} \\
\hline

BlockStart, mean $\pm$ SD &
$3.75\pm1.22$ ($n=12$) &
$3.23\pm1.36$ ($n=13$) &
$4.50\pm1.07$ ($n=8$) \\

PreTOR, mean $\pm$ SD &
$5.83\pm1.19$ ($n=12$) &
$5.92\pm1.26$ ($n=13$) &
$5.88\pm0.99$ ($n=8$) \\

Within-block change, mean $\pm$ SD &
$+2.08\pm1.56$ &
$+2.69\pm2.10$ &
$+1.38\pm1.41$ \\

\hline
\end{tabular}
\end{table*}

\end{document}